\documentclass[aps, pre, reprint, group edaddress, footinbib, showpacs,floatfix]{revtex4-1}

\usepackage{amsmath} 
\usepackage{graphicx}
\usepackage[caption=false]{subfig} 
\usepackage{dcolumn}
\usepackage{bm}
\usepackage{hyperref}
\usepackage[usenames]{xcolor}

\usepackage{verbatim}

\newcommand{\Zdchanged}{C_d}
\newcommand{\Zddchanged}{C_{dd}}
\newcommand{\Zdinstem}{S_d}
\newcommand{\Zddinstem}[2]{S_{dd}^{(#1 #2)}}
\newcommand{\Zddtwostem}{S_{dd}^{*}}

\newcommand{\Ztwoseg}{\tilde{Z}}
\newcommand{\ZtwosegV}[2]{\tilde{Z}(#1 ; #2)}
\newcommand{\Zdtwoseg}{\tilde{Z}_d}

\newcommand{\Ztwoseginstem}{\tilde{S}_d}
\newcommand{\ZtwoseginstemV}[3]{\tilde{S}_d(#1 ; #2 , #3)}
\newcommand{\Zdinstemenclosed}{\tilde{\tilde{S}}_d}

\newcommand{\ZddinstemMutual}{S_{dd,mutual}^{(11)}}

\newcommand{\ZddinstemOne}{S_{dd,1}^{(11)}}
\newcommand{\ZddinstemTwo}{S_{dd,2}^{(11)}}
\newcommand{\ZddinstemThree}{S_{dd,3}^{(11)}}

\begin{document}

\title{Interplay between single-stranded binding proteins on RNA secondary structure}
\author{Yi-Hsuan Lin}
\affiliation{Department of Physics, The Ohio State University, 191 W Woodruff
Av, Columbus, OH 43210-1107, USA}

\author{Ralf Bundschuh}
\affiliation{Department of Physics, Department of Chemistry \& Biochemistry,
Division of Hematology, Center for RNA Biology,
The Ohio State University, 191 W Woodruff
Av, Columbus, OH 43210-1107, USA}

\date{2013/05/09}

\begin{abstract}
RNA protein interactions control the fate of cellular RNAs and play an important role in gene regulation. An interdependency between such interactions allows for the implementation of logic functions in gene regulation. We investigate the interplay between RNA binding partners in the context of the statistical physics of RNA secondary structure, and define a linear correlation function between the two partners as a measurement of the interdependency of their binding events. We demonstrate the emergence of a long-range power-law behavior of this linear correlation function. This suggests RNA secondary structure driven interdependency between binding sites as a general mechanism for combinatorial post-transcriptional gene regulation.
\end{abstract}


\maketitle

\section{Introduction}

RNA is one of the fundamental biopolymers. It plays an important role in many biological functions~\cite{alberts,couzin02,riddihough05}. Each RNA molecule is a heteropolymer consisting of the four different nucleotides A, U, G and C in a specific order called the primary structure of the molecule. These nucleotides have a strong propensity to form Watson-Crick (i.e., G--C and A--U) base pairs. This pairing is achieved by the polymer bending back onto itself and thus forming intricate patterns of stems containing runs of base pairs stacked on top of each other connected by flexible linkers of unpaired nucleotides called a secondary structure~\cite{higgs00}. These structures then fold up into specific three-dimensional (tertiary) structures which enable some RNAs to perform catalytic functions while other RNAs mainly function as templates for transmitting genomic information.

After an RNA is transcribed in a cell, it is subject to a multitude of RNA processing, RNA localization, and RNA decay steps that together determine the fate of the molecule. These steps are regulated by interactions with RNA binding proteins or small RNAs such as microRNAs~\cite{mansfield09,morris10,pichon12}. Control over the fate of an RNA is known as post-transcriptional gene regulation and is an important component of information processing in a cell.  Information processing requires mechanisms that implement logical operations between inputs between various binding partners of an RNA or in physical terms mechanisms by which the binding of one partner influences the binding of another partner (in Biochemistry often denoted as ``cooperativity'' between binding sites). Considering the sizes of the binding partners and the distances between them, it has been speculated for long that the effective range of this interdependency of binding partners has to be much greater than the sizes of these binding partners, to integrate numerous binding partners on a long RNA molecule. However, the detailed mechanisms of this interdependency between RNA binding partners are still unclear.

Here, we propose a possible mechanism for this long-range interdependency between binding sites based on the RNA secondary structures (see Fig. \ref{fig:2nd-str}).  The main idea is that interactions between a single-stranded RNA binding protein or a microRNA and the RNA require that all bases of the specific target binding site are unpaired.  A successful RNA-protein binding event therefore excludes some of the permitted configurations from the originally protein-free RNA secondary structure ensemble. The whole ensemble of possible folding configurations thus has changed, and the probability of another protein or microRNA to bind on the same RNA at a different site will also change after the first successful binding. This leads to an RNA structure mediated interdependency between binding sites. 

In this paper, we consider only the case of two binding sites per RNA molecule, which is the simplest system to investigate this phenomenon in. We define a linear correlation function between the binding partners bound to an RNA molecule as the observable to quantify this interdependency, and investigate its properties with respect to the RNA structures. We find that this linear correlation function decays algebraically as a function of the distance between two protein binding sites, $D$. We discuss the linear correlation function for the homopolymer state in the molten phase of RNA secondary structures as well as for the heteropolymer state in the glass phase of RNA \cite{Pagnani00, Bundschuh02a, Hartmann01}. Such algebraically decaying correlation function provides long-range interactions between binding proteins or microRNAs on an RNA. Therefore, we show that long-range interdependency of binding sites that is necessary for the implementation of logical operations in post-transcriptional regulation is a generic property of RNA secondary structures and does not require any direct protein-protein interactions.

The paper is organized as follows. In Sec.~\ref{sec:reviewRNA2nd} we first give a brief review of RNA secondary structure, discussing its importance and constraints, and the reasons for focusing on it in our investigation. In Sec.~\ref{sec:simplifiedRNA} we introduce how we model protein RNA interactions. Then, we investigate the linear correlation function quantifying interdependency of binding sites in the simplest model for RNA secondary structure formation in Sec.~\ref{sec:sini_0}.  The absence of such protein binding correlations in the simplest model motivates us to include an important aspect of secondary structure formations, namely loop cost, into the model in Sec.~\ref{sec:sini_no0}. In that section, we establish that in the presence of a loop cost the linear correlation function between binding partners decays as a power-law with the distance between binding partners, symbolizing a long-range effect.  This central finding of our manuscript is supported through analytical calculations in the molten phase of RNA secondary structure and numerical calculations in the glass phase.  In Sec.~\ref{sec:loop_size_dependence} we show that this behavior is not altered when adding a size dependent loop penalty to the model.  Finally, we numerically establish a power-law behavior of the correlation function between RNA-binding proteins using the Vienna package~\cite{Hofacker94}, which represents the state of the art of quantitative modeling of realistic RNA secondary structures, in Sec.~\ref{sec:vienna} before concluding the manuscript.  Several of the more technical aspects are relegated to the appendices.


\section{Review of RNA secondary structure} \label{sec:reviewRNA2nd}

\begin{figure}
\includegraphics[angle=0,width=\columnwidth]{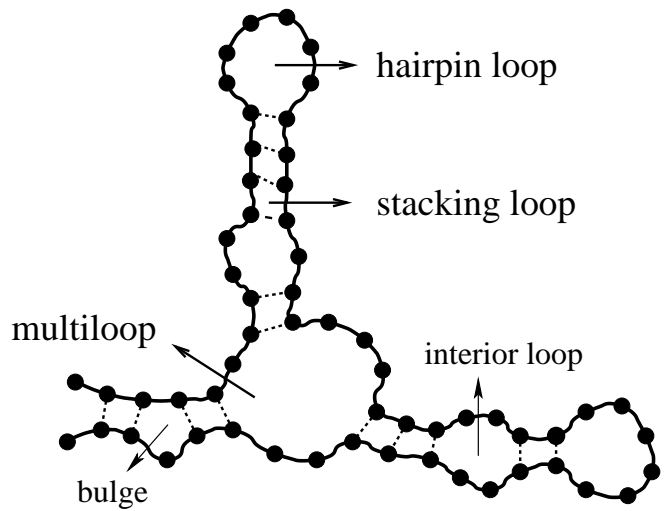}
\caption{A secondary structure of an RNA. The solid line shows the backbone of the RNA molecule. Dots are nucleotides, i.e. the bases of the RNA. Dashed lines are base-pair bonds. The regimes of multiple consecutive bonds are called stems, and the regions surrounded by unpaired bases are loops.  Depending on the number of stems eminating from the loops, they have different names as indicated in the figure.}
	\label{fig:2nd-str}
\end{figure}

\begin{figure}
\includegraphics[angle=0, width=\columnwidth]{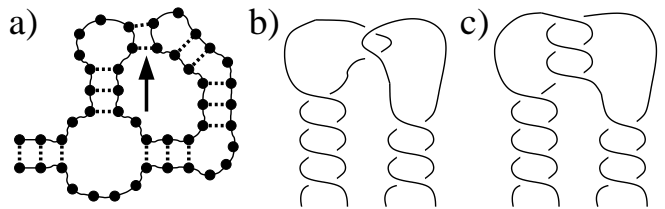}
\caption{(a) The base-pair bonds forming pseudoknots. Pseudoknots are excluded in the discussion of secondary structures because (b) short ones do not contribute much to the total binding energy, and (c) the longer ones are kinetically unlikely to form because of their complex double-helical structures.}
	\label{fig:pseudoknot}
\end{figure} 
 
A secondary structure of an RNA molecule describes its configuration as a base pairing pattern without specifying the three-dimensional arrangement of the molecule. In general, RNA molecules fold first into secondary structures which in turn fold into more complicated three-dimensional structures called tertiary structures. Moreover, even in the context of a higher-order structure, base pairing, i.e. the secondary structure, is the major contribution to the total folding energy. Thus, both from the dynamic and energetic point of view, secondary structure is the most important component of RNA folding. Therefore, it is appropriate to study secondary structures alone when trying to understand RNA folding phenomena; a long tradition in the field that we are also following here.

Each secondary structure is described by all of its formed base pairs, denoted as $(i,j)$ for a bond between the $i^{\text{th}}$ and $j^{\text{th}}$ nucleotide with $1 \leq i < j \leq N$ where $N$ is the sequence length. Different base pairs $(i,j)$ and $(i',j')$ are either independent ($i < j < i' < j'$) or nested ($i < i' < j' < j$). Pseudoknots, i.e. configurations with $i < i' < j < j' $, are usually excluded to make the structures more tractable both analytically and numerically. In practice, this exclusion is reasonable: long pesudoknots are kinetically forbidden (see Fig. \ref{fig:pseudoknot}), and short ones do not contribute much to the total binding energy. Moreover, because the RNA backbone is highly charged and pseudoknots increase the density of the molecule, their formation is relatively disfavored in low-salt conditions \cite{Bundschuh02a} and can even be ``turned off" in experiments to focus the study on the secondary structures \cite{Tinoco99}.  Thus, pseudoknots are commonly considered part of the tertiary structure of RNA and will be ignored in the rest of this paper.

In order to model RNA secondary structures, each structure $S$ needs to be assigned a folding energy $E[S]$. To calculate the energy $E[S]$ of a secondary structure $S$, it is necessary to clarify the contributions from different constituents of the structure. It is a good approximation to consider these contributions as local, i.e. to calculate the total energy $E[S]$ by summing over the contributions from each of the independent constituents, such as the stems and loops (see Fig. \ref{fig:2nd-str}). The dominant contribution from stems is the stacking energy, which is associated with two consecutive base-pair bonds and depends on the bases making up the stack. The contribution from a loop is more sophisticated. First, compared to a free chain, a loop of unpaired bases is entropically less favorable, and a free energy penalty, which depends on the loop length, has to be taken into account. Second, an enthalpy cost is involved, e.g., in bending the backbone for small loops and most importantly in the opening of a junction.  The combination of these two costs has been measured as a substantial length-independent term for loop initialization plus a relatively smaller length-dependent term~\cite{Mathews99}. In a complete model for real RNA, all these contributions depend not only on the size of a loop but also on its type (hairpin, interior, bulge, multiloop, see Fig.\ref{fig:2nd-str}) and sequence.

The partition function of an RNA sequence with $N$ nucleotides is calculated by considering all its possible secondary structures, summing over all of them as
\begin{equation}
Z(N) = \sum_{S \in \Omega (N)} e^{-\beta E[S]},
\end{equation}
where $\Omega(N)$ denotes the set of all possible secondary structures for the given sequence \cite{Higgs96, Bundschuh99, Pagnani00, Bundschuh02a}, and $\beta = 1/k_B T$ follows the traditional variable definition in statistical Physics. The set $\Omega(N)$ is not only constraint by the aforementioned secondary structure definition, but also constraint by the mechanical properties of the backbone of the RNA molecule. For a real RNA, the width of the double helix prohibits the formation of loops with less then three free base pairs. All secondary structures including these small loops have to be excluded from $\Omega(N)$. Considering all these sequence- and structure-dependent factors, the calculation of the partition function $Z(N)$ is complicated, requiring consideration of thousands of parameters, and therefore can only be handled numerically. In practice, the Vienna package~\cite{Hofacker94} implements such a complete model for the numerical calculation of thermodynamic properties of RNA molecules. 

For theoretical discussions, however, often simplified models are used to study generic properties of RNA folding. The most popular such model considers only the base-pair binding energy, instead of calculating the complicated stacking energy and loop cost. Thus, for each bond between two bases $i$ and $j$, the binding energy $\varepsilon_{ij}$ is calculated as
\begin{equation}  \label{eq:binding-eng_diff}
\varepsilon_{ij} = \left\{
\begin{aligned}
-u & \quad \text{if $(i,j)$ is an A-U or C-G pair} \\
u & \quad \text{otherwise},
\end{aligned}
\right.
\end{equation} 
and the total energy for a structure $S$ is just the sum of all binding energies, $E[S] = \sum_{(i,j) \in S} \varepsilon_{ij}$. The partition function can then be calculated by the recursive equation~\cite{deGennes68, Higgs96, Bundschuh99, Bundschuh02a}
\begin{equation}
Z_{(i,j)} = Z_{(i,j-1)} + \sum_{k=i}^{j-1} Z_{(i,k-1)} e^{-\beta \varepsilon_{kj}}Z_{(k+1,j-1)}, 
\end{equation}
where $Z_{(i,j)}$ denotes the partition function of the subsequence starting from the i$^{\mathrm{th}}$ and ending at the j$^{\mathrm{th}}$ nucleotide. The partition function of the whole sequence, $Z(N) \equiv Z_{(1,N)}$, can thus be calculated by iterating this recursion in $O(N^3)$ time. 

A second-order phase transition has been verified in this simplified model. At high enough temperature, the RNA molecule is in the so-called molten phase. In this phase, sequence dependence of the base-pair binding energy becomes irrelevant. Instead, the RNA molecule behaves after some coarse graining like a homopolymer, with an identical binding energy $\varepsilon_0$ for arbitrary pairs of nucleotides rendering its partition function analytically solvable~\cite{deGennes68}. In this case, $Z(N)$ is just a function of total sequence length, given as
\begin{equation} \label{eq:QN_simplest}
Z(N) \approx A_0(q) \frac{z_0^N(q)}{N^{3/2}}, 
\end{equation}
where $q \equiv \exp(-\beta \varepsilon_0)$, $z_0(q) = 1+2\sqrt{q}$, and $A_0(q) = [z_0^3(q)/(4\pi q^{3/2})]^{1/2}$ \cite{Bundschuh99}. As temperature decreases to the critical temperature, the phase transition occurs. The RNA transitions into the glass phase and becomes of noticeably heteropolymer nature, with sequence-dependent binding energy for each base-pair binding. The generic properties of glass phase RNA molecules are investigated by taking the quenched average of all random sequences, i.e. averaging over their free energies. These properties, including the occurrence of the phase transition itself, have been widely discussed numerically \cite{Bundschuh02a,Krzakala02}, and further verified by theoretical calculations on the basis of the renormalization group \cite{Lassig06, David07}.

A theoretical discussion is not necessary to be constraint to this simplest model. More complete but also complicated models can be constructed by adding the loop costs or even the stacking energy back into the model of the base-pair binding energy \cite{Muller03, Liu04}. Generally speaking, more detailed properties can be discovered by including more free energy terms and structure constraints in the model. However, the calculations also dramatically become much more difficult. To obtain a general idea of the behavior of the linear correlation function, here, instead of using the most detailed models from the start, we first restrict ourselves to the base-pair binding models, starting from the simplest one with only binding energy, and adding more free energy terms later if necessary. In the end we verify our findings using state of the art energy models including all the details necessary for quantitative RNA structure prediction.


\section{RNA-protein binding} \label{sec:simplifiedRNA}

In order to keep the language simpler, we will throughout the rest of this manuscript refer to the binding sites as protein binding sites, although they may represent microRNA binding sites as well. For the purpose of modeling RNA-protein binding events on RNA secondary structures, we consider merely the simplest and inevitable aspect: a bound protein, with a size $l$, would exclude all structures including any one of the $l$ base pairs in the footprint (i.e. the bases bound by the protein).  For all other bases we assume that they can form the same base pairs as without the protein binding~\cite{Forties10}. Even though, in practice, more sophisticated interactions, such as the excluded volume interaction between RNA and protein, occur around the footprint, we do not include them in our minimal model, for the purpose of a coarse-grained and conceptual investigation. 

All thermodynamic quantities concerning the RNA protein interactions can be derived from the partition function of the RNA-protein system,
\begin{equation}
Z = Z_0 + Z_1 e^{\beta \mu_1} + Z_2 e^{\beta \mu_2} + Z_{12} e^{\beta(\mu_1+\mu_2)}.
\end{equation}
In this expression $Z_0$ is the partition function over all secondary structures of the RNA without any protein binding; $Z_1$ and $Z_2$ are the limited partition functions, in which all bases at the first or second protein binding site are unpaired, respectively; $Z_{12}$ is the partition function for which all bases at both of the two protein binding sites are unpaired. $\mu_1$ and $\mu_2$ are the chemical potentials for the two proteins. Practically, the protein chemical potentials are controlled by their concentrations in solution as $\mu_k = \mu_{0,k} + k_B T \ln ( c_k/c_0)$, where $k = 1,2$, and $\mu_{0,1}, \mu_{0,2}$ and $c_0$ are characteristic parameters of the specific proteins, determined by experiments. The partition function can thus be rewritten as
\begin{equation}
Z = Z_0 + Z_1  \frac{c_1}{K_{d,1}^{(0)}} + Z_2 \frac{c_2}{K_{d,2}^{(0)}} + Z_{12} \frac{c_1 c_2}{K_{d,1}^{(0)} K_{d,2}^{(0)}},
\end{equation}
with $K_{d,k}^{(0)} = c_0 e^{-\beta \mu_{0,k}}$ the {\em bare} dissociation constant for a protein binding to an otherwise {\em unstructured} RNA. 

To quantify the interdependency between the protein binding sites, we introduce observables $P_k$ which are one if protein $k$ is bound and zero otherwise. If the the binding sites are independent, the thermodynamic average $\langle P_1 P_2 \rangle$ decouples into the product $\langle P_1 \rangle \langle P_2 \rangle$. Thus, we use $\langle P_1 P_2 \rangle - \langle P_1 \rangle \langle P_2 \rangle$ as a measure of interdependency of the binding sites. To discover generic characteristics of all nucleic acid sequences, we investigate the \textit{quenched average} of this protein-protein correlation function over all random RNA sequences,
\begin{equation}
G \equiv \overline{ \langle P_1 P_2 \rangle - \langle P_1 \rangle \langle P_2 \rangle}. 
\end{equation}
In practice, we will choose random RNA sequences in which each base is
chosen with equal probability from the four possibilities A, U, G, and
C, independently of the other bases. While structural RNAs have very
specific sequences that ensure their folding into a target structure,
the random sequence model is appropriate for messenger RNAs, the
sequences of which are not optimized for a specific structure and
which are anyways the more interesting targets for
post-transcriptional regulation. The linear correlation function is then
calculated as
\begin{equation} \label{eq:G12}
G = c_1 c_2 \frac{\partial^2 \overline{\ln Z} }{\partial c_1 \partial c_2} = \overline{\left( \frac{(Z_0 Z_{12}- Z_1 Z_2 )  c_1 c_2}{Z^2 K_{d,1}^{(0)} K_{d,2}^{(0)}}\right)}.
\end{equation}
In the following, we will investigate this linear correlation function as a function of the distance, $D$, between the two protein binding
sites.


\section{Simplest RNA folding model} \label{sec:sini_0}

\begin{figure}
\includegraphics[width=\columnwidth]{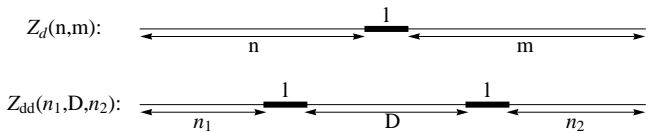}
\caption{In the molten phase, the base-pair binding energies of different nucleotides become identical. The limited partition functions with consideration of protein binding sites no longer depend on sequences and retard to functions of lengths of segments between binding sites.}
	\label{fig:ZddZdAll}
\end{figure}

As described in Sec. \ref{sec:reviewRNA2nd}, our investigation of the linear correlation function for protein binding sites starts from the simplest model, which includes only the base-pair binding energy defined in  Eq. (\ref{eq:binding-eng_diff}). Note that although we mentioned in Sec.~\ref{sec:reviewRNA2nd} that there is a minimum size for hairpin loops in real RNA, we do not impose such a constraint in this model for the purpose of a conceptual discussion. We first consider the high-temperature regime, where the RNA is in the molten phase, with the partition function given in Eq. (\ref{eq:QN_simplest}). Due to translational invariance, the limited partition functions also retard to functions of sequence length parameters $D$, $n_1$, $n_2$, and $l$, defined in Fig \ref{fig:ZddZdAll}, and can be written as
\begin{equation} \label{eq:ZddZdAll}
\begin{aligned}
& Z_1 = Z_d(n_1, D+l+n_2), \\
& Z_2 = Z_d(n_1+l+D, n_2), \\
& Z_{12} = Z_{dd}(n_1, D, n_2),
\end{aligned}
\end{equation}
where $Z_d$ and $Z_{dd}$ are limited molten phase partition functions with one and two stretches of $l$ unpaired bases, respectively (see Fig. \ref{fig:ZddZdAll}), and the lengths of the segments are constrained by the length of the whole molecule as $n_1+D+n_2+2l = N$. The exact form of $Z_d$ and $Z_{dd}$ can be derived by considering the following \textit{insertion} aspect: a limited partition function $Z_d(n,m)$ is constructed by inserting a blank segment, with the length equal to the footprint $l$, into the partition function $Z_0(n+m)$, between the $n^{\mathrm{th}}$ and $(n+1)^{\mathrm{st}}$ nucleotides. For the model including only pair binding energies, inserting such a blank segment does not affect the energies of any of the structures and thus Eq. (\ref{eq:ZddZdAll}) is further simplified to 
\begin{equation} 
\begin{aligned}
& Z_1 = Z_2 = Z_0(n_1+D+l+n_2) = Z_0(N-l), \\
& Z_{12} = Z_0(n_1+D+n_2) = Z_0(N-2l).
\end{aligned}
	\label{eq:Znm=Zn+m}
\end{equation}
The molten phase correlation function is then given by 
\begin{equation} 
\begin{aligned}
& g(D) \equiv G(D)\frac{K_{d,1}^{(0)} K_{d,2}^{(0)}}{c_1 c_2} \\
& = \frac{Z_0(N)Z_0(N-2l) - Z_0(N-l)^2}{Z^2(N,l,\{c_k/K_{d,k}^{(0)}\} )}, \\
\end{aligned}
	\label{eq:gD_simplest}
\end{equation}
completely independent of the distance between the protein binding sites and converges to $0$ as fast as $(l/N)^2$, as can be seen by substituting Eq. (\ref{eq:QN_simplest}) for all $Z_0$, yielding
\begin{equation}
g(D) = \frac{3 z_0^{-2l}}{2 [ (1+\frac{c_1}{z_0^l K_{d,1}^{(0)}}) (1+ \frac{c_2 }{z_0^{l}  K_{d,2}^{(0)}}) ]^2} \left( \frac{l}{N} \right)^2 + O\left(\! \frac{1}{N^3} \!\right). 
	\label{eq:gD_simplest_zero}
\end{equation}
We conclude that the model including only base pair bonds is {\em not} able to explain protein-binding correlations.

\section{RNA folding model with constant loop cost} \label{sec:sini_no0}

Since the simple energy model does not result in correlations between the binding sites, we are now going one step further than the simplest base-pair binding model and consider a loop cost. As discussed in Sec. \ref{sec:reviewRNA2nd}, the complete form of the loop cost includes a constant term for loop initialization and a length-dependent term for extension. Since the constant term is generally much greater than the length-dependent term, it is appropriate to take loops into account only through this constant term. Although this loop cost has in principle entropic and enthalpic components, we follow the literature~\cite{Mathews99} and model it when varying temperature as purely entropic, i.e. take loops into account through a temperature-independent Boltzmann factor, $\exp(-s_0) < 1$. We note, however, that this choice only affects the detailed temperature dependence and not the presence of protein-protein correlations per se.

In oder to compute the partition function for an RNA folding model with the pairing energies given in Eq. (\ref{eq:binding-eng_diff}) and constant loop cost $s_0$, two different auxiliary partition functions are required. They are $Z_{b (i,j)}$, the partition function for structures on a substrand starting at the i$^{th}$ nucleotide and ending at the j$^{th}$ one, having a bond between the first and last nucleotides, and $Z_{(i,j)}$, the partition function starting from the i$^{th}$ nucleotide and ending at the j$^{th}$ one without any further constraints. These quantities obey the recursion equations~\cite{deGennes68,Nussinov78,Waterman78,McCaskill90}
\begin{equation}
\begin{aligned}
Z_{b(i,j)} = & q_{ij} \left[Z_{b(i\!+\!1,j\!-\!1)} + \xi(Z_{(i+1,j\!-\!1)} - Z_{b(i\!+\!1,j\!-\!1)}) \right] \, \text{and} \\
Z_{(i,j)} = & Z_{(i,j-1)} + \sum_{m=i}^{j-1} Z_{(i,m-1)} Z_{b(m,j)},
\end{aligned}
	\label{eq:recursion}
\end{equation}
where $\xi \equiv \exp(-s_0)$ and  $q_{ij} \equiv \exp( \beta \varepsilon_{ij})$, which can be iterated numerically to compute the full partition function $Z_{(1,N)}$ for arbitrary sequences and loop penalties $\xi$ in $O(N^3)$ time.  Again, we neglect constraints on the size of hairpin loops for simplicity.


\subsection{Molten phase} 

Similar to the simplest model of only base-pair binding energy, the
recursion in Eq.~(\ref{eq:recursion}) can be analytically solved for a
molten-phase RNA with loop cost by substituting a homogeneous $q$ for
all $q_{ij}$, yielding (see~\cite{Muller03,Liu04,Tamm07} and Appendix \ref{app:ZN})
\begin{equation}
Z_0(N) = A(q,\xi) \frac{z^{N}(q,\xi)}{N^{3/2}}[1+O(N^{-1})],
	 \label{eq:G_BC}
\end{equation}
where $A$ and $z$ are non-universal parameters depending on the values
of $q$ and $\xi$. Comparing Eqs. (\ref{eq:G_BC}) and
(\ref{eq:QN_simplest}), the partition function with and without the
entropy cost have exactly the same form --- the loop cost is
irrelevant for the asymptotic behavior as also known in the context of
DNA melting~\cite{Grosberg94} for a long time.

\begin{figure}
\includegraphics[width=1\columnwidth]{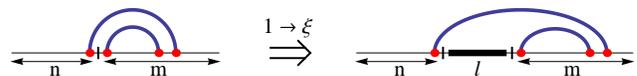} 
\caption{Structures in which the footprint is in a stem, i.e., inserted between two consecutive base-pair bonds. Such structures contribute differently in $Z_0(n+m)$ (left) and $Z_d(n,m)$ (right). }
	\label{fig:Zdstem} 
\end{figure}


However, for the limited partition functions $Z_d$ and $Z_{dd}$, the
relationships corresponding to Eqs.~(\ref{eq:Znm=Zn+m}) no longer hold
when the loop cost is non-zero, i.e. $Z_d(n,m) \neq Z_0(n+m)$.  This
can be seen as follows: Once an unpaired segment of length $l$ is
inserted into $Z_0(n+m)$, it can either create a new loop, or extend
an existing loop. The secondary structures taken into account by
$Z_0(n+m)$ are thus separated into two groups, reacting differently to
the insertion. If a structure of $Z_0(n+m)$ has a stem cross the
$n^{th}$ and $(n+1)^{st}$ base pairs, the insertion of the footprint
between the two nucleotides {\em changes} the contribution of this
structure to $Z_d(n,m)$ by a loop factor $\xi$ since the insertion
creates a new loop as shown in Fig. \ref{fig:Zdstem}. For all other
structures, the contribution to $Z_0(n+m)$ and $Z_d(n,m)$ are the
same. It is therefore necessary to distinguish which structure belongs
to which one of the two groups. Defining the partition function for
all structures with a stem containing the $n^{th}$ and $(n+1)^{st}$
base pairs (i.e., the partition function for all structures for which
the insertion of a footprint generates a new loop) as
$\Zdinstem(n,m)$, the limited partition function can be expressed as
\begin{equation} 
\begin{aligned}
Z_d(n,m) & = Z_0(n+m) - (1-\xi) \Zdinstem(n,m). \\
& \equiv Z_0(n+m) + \Zdchanged(n,m)
	\label{eq:Zd-S-Z0}
\end{aligned}
\end{equation}
Notice that $\Zdchanged(n,m)=-(1-\xi)\Zdinstem(n,m)$ is calculated as
the contribution to $Z_0(n+m)$ in the absence of the inserted
footprint. The first term $Z_0(n+m)$ is just the one used in the
simplest model --- albeit itself dependent on $\xi$ as given by
Eq.~(\ref{eq:G_BC}) --- and the subsequent term, proportional to
$(1-\xi)$, is the effect of the footprint insertion resulting from a
nonzero loop cost.

A similar strategy as Eq. (\ref{eq:Zd-S-Z0}) can also be applied to the calculation of $Z_{dd}$ by defining its changed term via
\begin{equation}
Z_{dd}(n_1,D,n_2) = Z_0(n_1+D+n_2) + \Zddchanged(n_1,D,n_2).
\end{equation}
However, since now there are two footprints of the protein,
$\Zddchanged$ is more complicated than $\Zdchanged$ and cannot be
simply written down as a term proportional to $(1-\xi)$. Considering
that each insertion of an unpaired stretch of bases into a
base stack of a stem (no matter how many stretches are inserted into a
stack) changes the contribution of the structure by a
factor of $\xi$, four different configurations, affected differently
by the insertions, are included in $\Zddchanged$:

(i) The first footprint is in a stem but the second one is not,
contributing a change of $(\xi-1)$.

(ii) The second footprint is in a stem but the first one is not,
contributing a change of $(\xi-1)$.

(iii) Both footprints are in different stems or different base pair
stacks of the same stem, contributing a change of $(\xi^2-1)$.

(iv) Both footprints are on opposite sides of the same base stack of a
stem, contributing a change of $(\xi-1)$.

We describe the partition functions for the first three configurations
with the help of the partition functions $\Zddinstem{a}{b}$, where the
labels $a$ and $b$ describe constraints on the first and the second
binding site respectively. A label of $1$ for $a$ or $b$ denotes that
the corresponding binding site is in a stack, a label of $0$ denotes
that the binding site is not in a stack, and a label of $\times$
indicates that there is no constraint for the corresponding binding
site.  In addition, we introduce the partition function for all
configurations in which the locations of both footprints are on
opposite sides of the same stack as $\Zddtwostem$.  The changed term
$C_{dd}$ can then be written as
\begin{equation}
\begin{aligned}
& C_{dd}(n_1,D,n_2) \\
= & (\xi-1)\Zddinstem{1}{0} + (\xi-1)\Zddinstem{0}{1} \\
& \qquad \qquad +  (\xi^2-1)(\Zddinstem{1}{1} - \Zddtwostem) + (\xi-1) \Zddtwostem \\
= & (\xi-1)(\Zddinstem{1}{\times} - \Zddinstem{1}{1}) + (\xi-1)( \Zddinstem{\times}{1} - \Zddinstem{1}{1}) \\
& \qquad \qquad+  (\xi^2-1)(\Zddinstem{1}{1} - \Zddtwostem ) + (\xi-1)\Zddtwostem \\
= & (\xi-1)(\Zddinstem{1}{\times} + \Zddinstem{\times}{1}) + (\xi-1)^2\Zddinstem{1}{1}+ \xi(1-\xi) \Zddtwostem. 
\end{aligned}
\end{equation}
where we omit the arguments $(n_1,D,n_2)$ for each $\Zddinstem{a}{b}$
and $\Zddtwostem$ for the sake of clarity.  We note that the limited
partition functions $\Zddinstem{1}{\times}$ and $\Zddinstem{\times}{1}$,
since they only contain constraints at the location of one of the
footprints, can be exactly expressed as
\begin{equation}
\begin{aligned}
\Zddinstem{1}{\times}(n_1,D,n_2) & = \Zdinstem(n_1,D+n_2) \; \mathrm{, \; and}\\
\Zddinstem{\times}{1}(n_1,D,n_2) & = \Zdinstem(n_1+D,n_2) \\
\end{aligned}
\end{equation}
The limited partition function $Z_{dd}$ is thus given as
\begin{equation}
\begin{aligned}
& Z_{dd}(n_1,D,n_2) \\
& = Z_0(n_1+D+n_2) \\
& - \quad (1-\xi)\left[\Zdinstem(n_1,D+n_2)+ \Zdinstem(n_1+D,n_2) \right] \\
& + \quad (1-\xi)^2\Zddinstem{1}{1}(n_1,D,n_2) + \xi(1-\xi)  \Zddtwostem (n_1,D,n_2).
\end{aligned}
	\label{eq:Zdd_final}
\end{equation}

\begin{figure}
\includegraphics[width=\columnwidth]{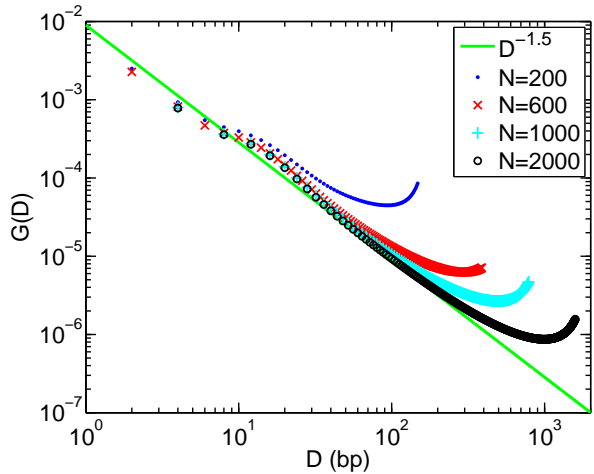}
\caption{Numerical calculation of the protein protein correlation $g(D)$ on RNA polymers of $N=200,600,1000,$ and $2000$ based on the molten-phase homopolymer model, with $q = \exp(0.2)$, $s_0 = 5$ and $c_i/K_{d,i}^{(0)} = 0$. The footprint is set to be $l = 6$. The correlation satisfy a power law $g(D) \sim D^{-3/2}$ (solid line) in the regime $N \gg D \gg l$.}
	\label{fig:g-molten}
\end{figure}

At this point, we have investigated the effect of protein binding site
insertions on all the limited partition functions required to
calculate the molten phase correlation function $g(D)$ for this model
with constant loop cost.  Unfortunately, the quantities $\Zdinstem$,
$\Zddinstem{1}{1}$ and $\Zddtwostem$, can not be calculated exactly
and we have to make two approximations.  First, we consider the limit
$N/2 \approx n_1 \approx n_2 \gg D \gg l$, where $N = n_1+D+n_2+2l$ is
the length of the whole RNA molecule.  Second, we investigate the
correlation function $g(D)$ only perturbatively in the loop cost or
more precisely as an expansion in $(1-\xi)$.  This investigation, the
details of which are given in the appendices, shows that already an
infinitesimally small loop cost will yield a non-zero correlation
function that shows a power law dependence on $D$.  We will later
demonstrate numerically (see Fig.~\ref{fig:g-molten}) that this result
remains valid in the biologically relevant regime of small $\xi$ (or
$(1-\xi)\approx1$).

More specifically, expanding the limited partition functions to the
appropriate orders of $(1-\xi)$ and then inserting these expansions
into the definition of the correlation function $g(D)$ as shown in
Appendices~\ref{app:gD_calculation} and~\ref{app:second_order} yields
\begin{widetext}
\begin{equation} 
\begin{aligned}
g(D) & = \frac{Z_0(N) Z_{dd}(n_1, D , n_2) - Z_d(n_1, D+l+n_2)Z_d(n_1+l+D, n_2)}{Z(N,D,l, \{c_k/K_{d,k}^{(0)}\})^2} \\
& =(1-\xi) \frac{\mathcal{A}(q,\xi,\{c_k/K_{d,k}^{(0)}\})}{D^{3/2}} + O\left(\frac{1}{N} \right),
\end{aligned}
	\label{eq:g12-molten}
\end{equation}
\end{widetext}
in the limit of $N \gg D \gg l$, where $N$ is sequence length, $D$ is
distance between protein binding sites, and $l$ is the footprint of
protein binding. The prefactor $\mathcal{A}$ converges to $A(q,\xi)
z(q,\xi) ^{-2(l+2)} q^2 \xi^2$ in the limit $c_k \ll
K_{d,k}^{(0)}$ and $1-\xi \ll 1$, and has to be determined numerically
in the general case. The leading order term, which obeys a power law
with an exponent of $3/2$, is the zeroth order term of $1/N$ and thus
does not vanish as $N \rightarrow \infty$.

As we have pointed out above, the analytical result that the
correlation function $g(D)$ obeys the power law is only a perturbative
result for small $(1-\xi)$, i.e., small loop costs.  It is somewhat
reassuring that the first and second order terms in this expansion
show the same power law.  Nevertheless, loop costs in real RNA tend to
be large which is why in Fig. \ref{fig:g-molten} we numerically
verify the occurrence of the same power-law $g(D)$ in the molten phase
model with a substantially large loop cost, $s_0 = 5$. The combination
of the perturbative calculation and the numerical
evidence confirms that any non-zero loop cost contributing to the
partition function as $\xi \neq 1$ qualitatively changes the
properties of the protein-protein correlation function leading to a
$D^{-3/2}$ long-range correlation between the multiple binding
partners.

\subsection{Glass phase}

In reality, it is believed that RNA molecules at room temperature are not in the molten but rather in the so-called glass phase~\cite{Pagnani00,Hartmann01,Bundschuh02a}. Thus, it is necessary to investigate the correlation function $G(D)$ in that phase. When temperature decreases, the differences between the $\varepsilon_{ij}$ become relevant, and the disorder breaks the homopolymer assumption for RNA molecules. Therefore, the limited partition functions can no longer be appropriately expressed in terms of the translationally invariant $Z_d$ and $Z_{dd}$, and Eq. (\ref{eq:g12-molten}) is not applicable, either. Moreover, the effect of the denominator in Eq. (\ref{eq:G12}) and thus of the protein-binding parameters $c_k/K_{d,k}^{(0)}$ is now uncertain. These parameters may now play an important role in the form of the correlation function, instead of only modifying the prefactor as in Eqs. (\ref{eq:gD_simplest}) and (\ref{eq:g12-molten}).

To clarify the effect of protein binding parameters, we consider the ratio $Z_0/Z_k$, where $k = 1, 2$, and compare its value with the corresponding protein-binding term $c_k/K_{d,k}^{(0)}$, to quantify the protein concentrations. We thus introduce the \textit{effective} dissociation constant for each individual RNA sequence by taking into account the free energy difference between free and bound RNAs, 
\begin{equation}
\Delta F_k \equiv -k_BT \ln(Z_0/Z_k).
\end{equation}
The generic effective dissociation constant for all random RNA sequences is then derived by considering the quenched average of the free energy difference, $\overline{\Delta F_k}$, yielding
\begin{equation}
K_{d,k} = K_{d,k}^{(0)} e^{-\beta \overline{\Delta F_k}}
\end{equation}
Using this concept, we consider the concentration dependence of $G(D)$ in the following three different regimes: (i) dilute concentration, where $c_k \ll K_{d,k}$, (ii) saturated concentration, where $c_k \gg K_{d,k}$, and (iii) normal concentration, where $ c_k \approx K_{d,k}$. In the dilute regime, proteins essentially never bind to the RNA while in the saturated regime they remain nearly always bound. Thus, most biochemical reactions occur in regime (iii).

\begin{figure}
\includegraphics[width=\columnwidth]{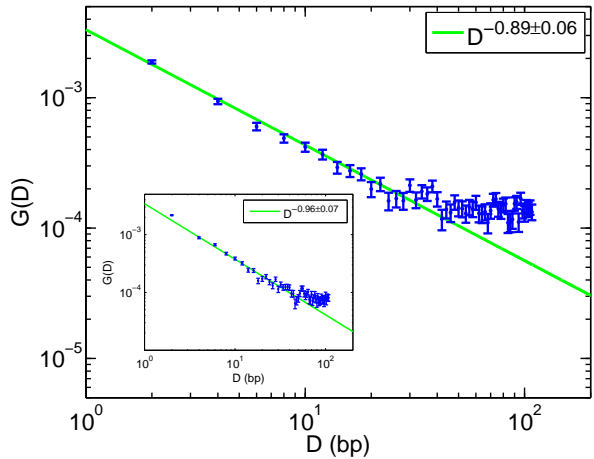}
\caption{Numerical calculation of the correlation function $G(D)$ on RNA polymers of 200 nucleotides for normal protein concentrations at low temperature, $k_BT = 0.3u$. The loop cost is $s_0 = 5$. The quenched average is taken over 400,000 random sequences. Footprints are $l = 6$ in the main graph and $l = 1$ in the inset. 
Protein binding parameters are $K_{d,1}^{(0)} = K_{d,2}^{(0)} = 0.1$nM. Concentrations are $c_1 = c_2 = 100$nM, yielding $c_k/K_{d,k} \approx 3.84$ by $e^{\beta \overline{\Delta F_k}} \approx 0.00384$ for both $k = 1,2$ in main graph, and $c_1 = c_2 = 2$nM to have $c_k/K_{d,k} \approx 3.72$ by $e^{\beta \overline{\Delta F_k}} \approx 0.186$ in the inset. The correlation functions of the two graphs follow power laws as $G(D) \sim 1/D^{0.9}$.}
	\label{fig:logCor_Own}
\end{figure}

In Fig.~\ref{fig:logCor_Own} we show the correlation function of the simplified model using base pairing energies from Eq.~(\ref{eq:binding-eng_diff}), a loop cost of $s_0=5$, and random sequences with equal probabilities for the four possible nucleotides for normal protein concentrations ($c_k/K_{d,k} \approx 4$). A similar calculation resulting from a shorter footprint is shown in the inset ($l = 1$ compared to the $l = 6$ in the main graph) to allow an evaluation of the effect of footprint size.  Again, we observe a power law dependence of the correlation function on the distance between the binding sites.  We verified this power law dependence in the normal protein concentration regime numerically for a whole range of temperatures $k_BT/u = 0.1 \sim 0.9$ (which spans both sides of the glass transition as determined by a peak in the specific heat in the vicinity of $k_BT\approx0.3u$, data not shown).

\begin{figure}
$\begin{array}{cc}
\subfloat[]{\includegraphics[width=.45\columnwidth]{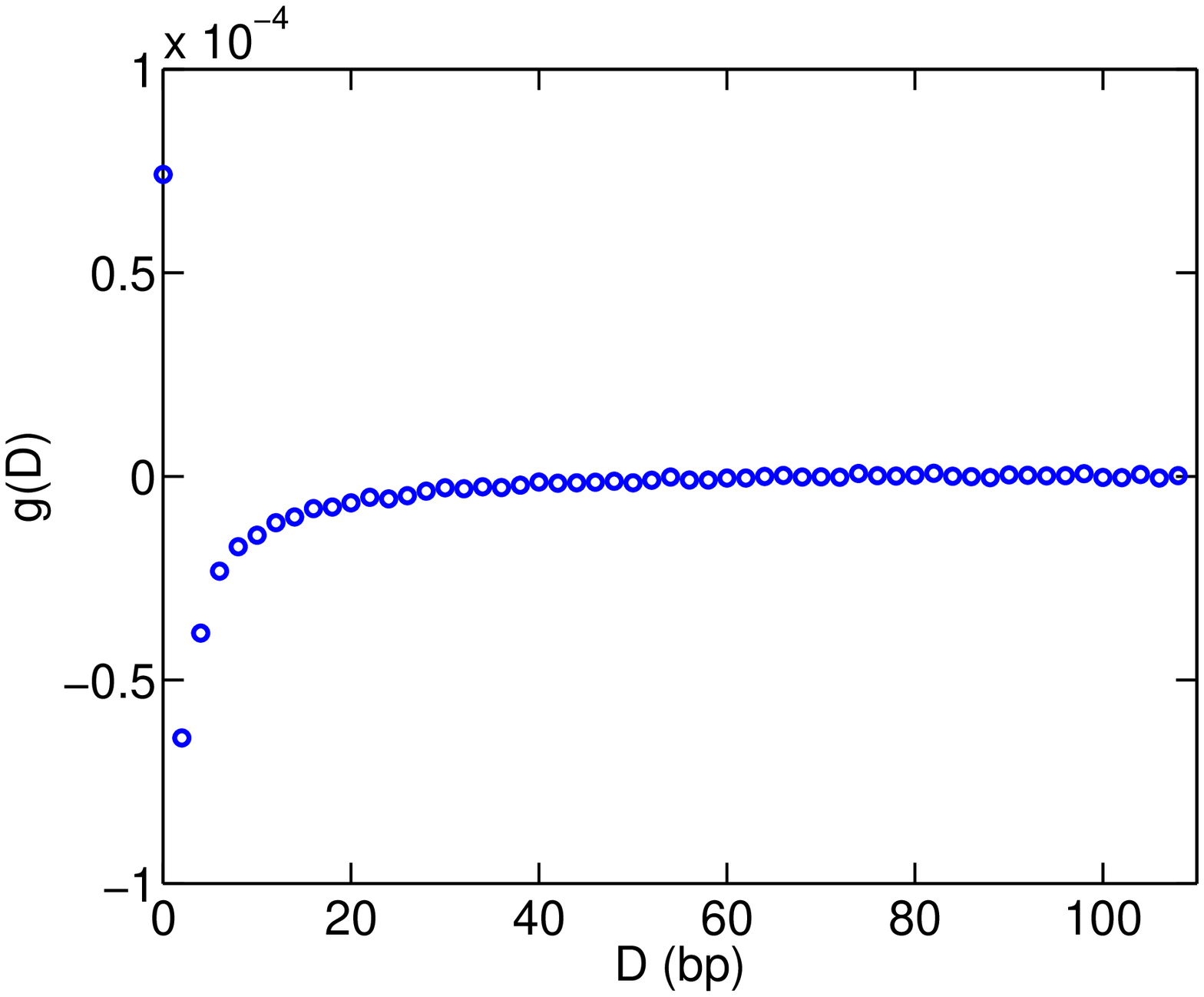} \label{fig:Own_Dil} } &
\subfloat[]{\includegraphics[width=.45\columnwidth]{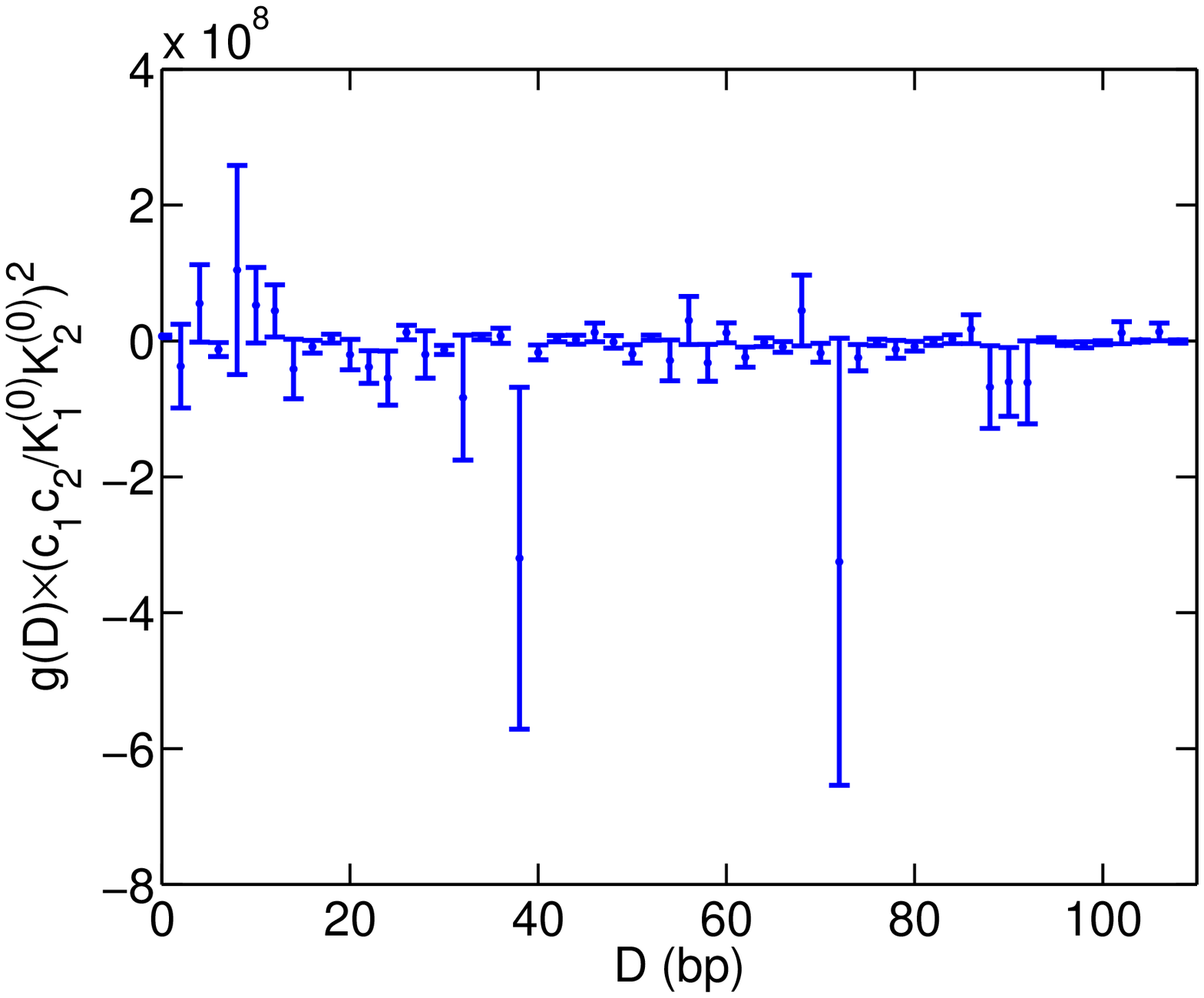} \label{fig:Own_Sat} }
\end{array}$
\caption{Numerical calculation of the protein-protein correlation g(D) on model RNA polymers of 200 nucleotides for (a) extremely dilute and (b) extremely saturated protein concentrations. The quenched average is taken over 400,000 random sequences.  Temperature is set to $k_BT=0.3u$ and the secondary structure parameters are the same as those of Fig.~\protect\ref{fig:logCor_Own}. No power law dependence of the correlation function appears.}
	\label{fig:Own_SatAndDil}
\end{figure}

However, in contrary to the molten phase results in the last subsection, it turns out that protein concentration is an important factor in the glass phase.  Fig.~\ref{fig:Own_SatAndDil} shows the correlation function for the same model and parameters as above but for protein concentrations in the dilute and the saturated regime, respectively.  In this case, the correlation function does not show a power law behavior.  These discoveries suggest that while the power-law correlation is sensitive to temperature and protein binding parameters, it generically occurs precisely in the biologically relevant regime where $c_k \approx K_{d,k}$.

\section{RNA folding model with loop-size dependent cost}\label{sec:loop_size_dependence}

In Sec.~\ref{sec:reviewRNA2nd} we explained that the loop cost comprises initialization and extension components. In practice, this extension penalty is due to the entropic loss upon forming a closed loop, and is thus proportional to $\ln(L)$ with $L$ the loop length \cite{Muller03}. Taking this logarithmic dependence into account numerically results in an algorithm for RNA partition functions of complexity $O(N^4)$, much less efficient then the aforementioned $O(N^3)$ one. Thus, current state of the art RNA structure prediction tools such as MFOLD~\cite{Zuker03} and the Vienna package~\cite{Hofacker94} linearize the loop cost for multiloops and interior loops by approximating $\ln(L) \approx L-1$, supposing that loops with very large $L$, where the difference between the logarithmic and linear dependence becomes noticeable, are extremely energetically unfavorable and thus forbidden.

Thus, we consider here the effect of such a linear loop cost. The total Boltzmann factor for a loop with $L$ free bases now becomes $\exp(-s_0 - \nu L)$ with $\nu$ the extension penalty per free base in a loop. To take into account the effect of the extension penalty, one more auxiliary partition function, $Z_{m,(i,j)}$ for a substrand from the $i^{th}$ to the $j^{th}$ base in the context of a closed base pair $(i',j')$ with $i' < i <j <j'$ is required in addition to the partition functions $Z_{(i,j)}$ and $Z_{b(i,j)}$ used in Eq~(\ref{eq:recursion}).  The new recursive equations for calculating the three partition functions are then given by
\begin{equation}
\begin{aligned}
Z_{b(i,j)} = & q_{ij} \left[Z_{b(i+1,j-1)} + \xi(Z_{m(i+1,j-1)} \!- \! Z_{b(i+1,j-1)}) \right], \\
Z_{m(i,j)} = & \tilde{\xi}  Z_{m(i,j-1)} + \sum_{k=i}^{j-1} Z_{m(i,k-1)} Z_{b(k,j)}, \\
Z_{(i,j)} = & Z_{(i,j-1)} + \sum_{k=i}^{j-1} Z_{(i,k-1)} Z_{b(k,j)}, \\
	\label{eq:recursion_Sext}
\end{aligned}
\end{equation}
with $\tilde{\xi} \equiv \exp(-\nu)$.  Again, we neglect constraints on the size of hairpin loops for simplicity.

The extension loop cost results in a new property different from those in the aforementioned two RNA folding models, yielding very different asymptotic RNA secondary structure partition functions.  This new property becomes apparent by adding a free energy contribution of $k_BT\nu$ to every base, which amounts to a physically irrelevant overall shift in all free energies by $Nk_BT\nu$.  For paired bases this amounts to shifting the pairing energy to $u' = u + 2 k_B T \nu$. For all bases inside of a closed bond (i.e. the ones included in the partition function $Z_m$), this transformation just retards the model back to the one including only a constant loop cost; for the unpaired bases \textit{outside of} any bond, however, it results in an additional {\em gain} in free energy of $k_BT \nu$. Such a gain in free energy on the free bases can be viewed as the work resulting from a constant force stretching the RNA molecule, $k_B T \nu = f \cdot x$, thus mapping an RNA molecule with extension loop cost to a molecule with constant loop cost under tension. 

A second order phase transition has been discovered in such RNA molecules under tension~\cite{Muller03, Montanari01, Muller02}.  In the homopolymer case, as the stretching force is weak, the asymptotic RNA partition function still holds the form $Z_0(N) \approx A z^N/N^{3/2}$. Once the force crosses a threshold, the phase transition occurs, and the asymptotic partition function becomes $Z_0(N) \approx A' z^N$, a purely exponentially increasing function of sequence length \cite{Montanari01, Muller02}. In our model with extension loop cost, the phase transition is effectively driven by tuning the temperature and thus tuning the parameter $u'/k_B T - 2\nu$.  At high temperatures, the $\nu$-term is dominant and thus the RNA molecule is in the stretched phase characterized by the purely exponential partition function.  Below a critical temperature, the $\nu$-term is irrelevant and the molecule is in the regular molten or glass phase.

\begin{figure}
\includegraphics[width=\columnwidth]{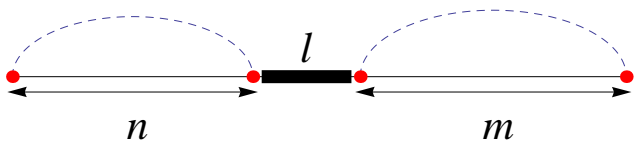} \label{fig:Zdnaked} 
\caption{A structure in which the footprint is naked. Dashed lines represent the two independent substrands in the calculation for partition functions. Such structures contribute identically in $Z_d(n,m)$ and $Z_0(n+m)$.}
	\label{fig:Zd_naked}
\end{figure}

If the parameters are chosen such that the molecule is below the critical temperature of the stretching transition but above the glass transition temperature, i.e. where $Z_0(N) \approx A z^N/N^{3/2}$, we can again calculate the protein-protein correlation function.  In fact, this calculation is very similar to the one for the RNA folding model including a constant loop cost. For most structures, their contributions to the partition function $Z_d(n,m)$ (where a protein binding footprint is to be inserted after the $n^{th}$ base pair) are simply the ones for $Z_0(n+m)$ multiplied by a penalty factor $\tilde{\xi}^l$ or $\xi \tilde{\xi}^l$, since either the inserted $l$-base long footprint extends an existing loop or creates a new loop, respectively.  Considering these two types of configurations, the limited partition function $Z_d$ for a model with extension loop cost is simply the one for the model with constant loop cost multiplied by $\tilde{\xi}^{l}$ (with appropriate parameters $A(q, \xi, \tilde{\xi})$ and $z(q, \xi, \tilde{\xi})$).  There are some structures, in which the footprint neither creates a new loop nor extends an existing loop but is ``naked" (see Fig. \ref{fig:Zd_naked}). These structures contribute identically in $Z_d(n,m)$ and $Z_0(n+m)$, i.e., without an additional factor $\tilde{\xi}^l$.  However, the partition function for these naked structures is simply $Z_0(n)Z_0(m)=O(z^N N^{-3})$, which can be ignored in the limit of large $N/2\approx n\approx m$ compared to the leading order term in $Z_d$ of order $O(z^N N^{-3/2})$ as we have seen numerous time in the calculations for the model with constant loop cost. Similarly, in $Z_{dd}$ (the partition function into which two footprints of length $l$ are to be inserted), the limited partition functions for structures with at least one naked footprint are all at the order of $O(z^N N^{-3})$ and thus negligible, leading to a $Z_{dd}$ identical to the one for the model including constant loop cost multiplied by $\tilde{\xi}^{2l}$. 

In consequence, for a model including a linearized extension loop
cost, once the asymptotic partition function holds the form $Z_0(N)
\approx A z^N/N^{3/2}$, the asymptotic limited partition functions
$Z_d$ and $Z_{dd}$ are identical to those in the model including only
a constant loop cost up to factors $\tilde{\xi}^l$ and
$\tilde{\xi}^{2l}$, respectively, and thus the protein-protein
correlation function has the same form as that in
Eq. (\ref{eq:g12-molten}), with only additional factors of $\tilde{\xi}^l$
multiplying the protein concentrations.

\begin{figure}
\includegraphics[width=\columnwidth]{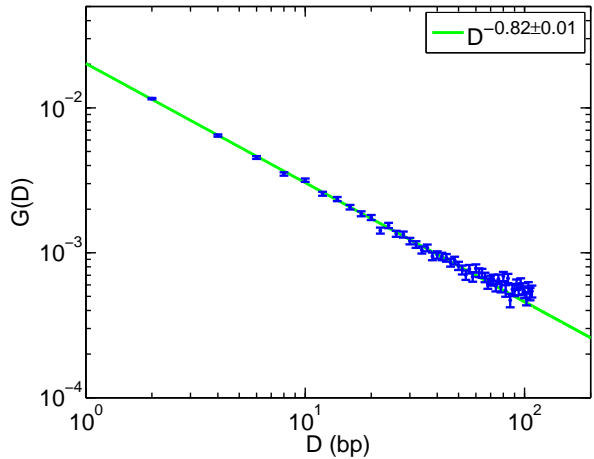}
\caption{Numerical calculation of correlation function $G(D)$ on RNA polymers of 200 nucleotides for normal protein concentrations at low temperature, $k_BT = 0.3u$. The loop cost is $s_0 = 5$ and $\nu = 0.5$. The quenched average is taken over 100,000 random sequences. Protein binding parameters are $K_{d,1}^{(0)} = K_{d,2}^{(0)} = 0.1$nM and $c_1 = c_2 = 100$nM, with $e^{\beta \overline{\Delta F_k}} \approx 0.000401$ for both $k = 1,2$. The footprint is equal to $l=6$ base pairs. The correlation function follows a power law as $G(D) \sim 1/D^{0.8}$.}
	\label{fig:logCor_Own_Sext}
\end{figure}

Since in practice the extension cost per free base in a loop, $\nu$, is much less than the loop initialization cost $s_0$, the critical temperature of the force-driven transition is typically high enough such that the force-driven phase transition occurs earlier than the glass-molten phase transition as temperature decreases. The properties of the glass phase (limited) partition functions are thus believed not to be affected by the stretched phase and thus to be similar to those of the model including only a constant loop cost.  Therefore, we also expect a similar behavior of the correlation function to that in the model with constant loop cost. This expectation is verified numerically in Fig.~\ref{fig:logCor_Own_Sext}, where the correlation function between two protein binding sites, again, decays as a power law of the distance between the sites in the normal concentration regime. It is worth pointing out that the quenched average over random sequences for the correlation of the model with loop extension penalty converges faster than that of  the model with only constant loop penalty and yields an even more convincing power law.  This might imply that, in the glass phase, for the model with extension penalty, the energy distribution of secondary structures has a narrower peak around the global maximum; for the model with only constant penalty, however, the energy distribution is wider, and thus the average has to be taken over much more random sequences for convergence. The relationship between the energy landscape in the glass phase and the loop penalty as a function of loop length would be a valuable topic for more discussion in the future.

\section{Realistic RNA folding model}\label{sec:vienna}

In the previous sections, we have established that the protein-protein
correlation function is a power law in serveral simplified models of
RNA folding as long as they include a loop cost.  While simplified
models are useful to disentangle the mechanism of and the minimum
requirements for the power law behavior, an obvious question is if
this finding is an artifact of the simplified model or if it applies
to real RNA molecules as well.  To this end, we numerically study the
same protein-protein correlation function as before using the Vienna
package~\cite{Hofacker94}, which represents the state of the art in
quantitative RNA secondary structure prediction using thousands of
measured free energy parameters including the very important stacking
free energies.  We apply the constraint folding capabilities of the
Vienna package to exclude the footprints (i.e., the bases bound by the
protein) from participating in base pairing, which allows us to
calculate the limited partition functions $Z_1$, $Z_2$, and $Z_{12}$
for arbitrary RNA sequences and distances between the protein binding
sites.

\begin{figure}
\includegraphics[width=\columnwidth]{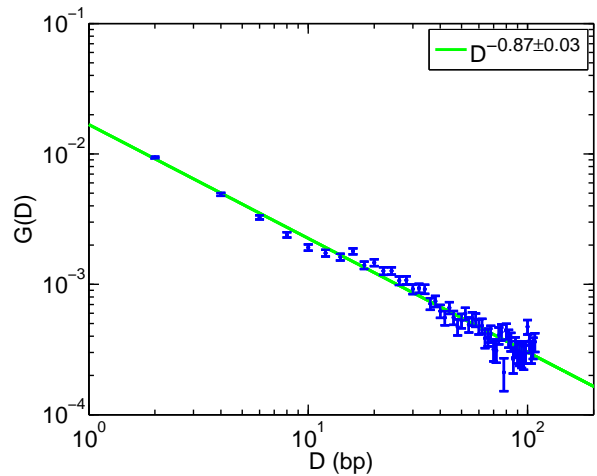}
\caption{Numerical calculation of protein-protein correlation $G$ on RNA polymers of 200 nucleotides. The calculation is based on the Vienna package and the quenched average is taken over 100,000 random sequences. Temperature is set to $T = 37^{\circ}\rm{C}$. The parameters for protein binding are $c_1/K_{d,1}^{(0)} = c_2/K_{d,2}^{(0)} = 100$ with $e^{\beta \overline{\Delta F_k}} \approx 0.00410$, which can be easily achieved by RNA binding proteins and are chosen to allow the proteins to compete with the RNA base pairing without completely outcompeting it. The length of the binding sites (footprint) is equal to 6 bases. Distance is expressed by the number of base pairs (bp) between the two binding sites. This log-log plot shows a regression to a power-law decay function $G(D) \sim 1/D^{0.9}$. } 
	\label{fig:logCor_Vienna}
\end{figure}

Based on our experience with the simplified models, we choose the
protein concentrations and binding constants in the biologically
reasonable regime (iii) where the binding sites are neither
essentially empty nor completely saturated.  We average over $100{,}000$
random sequences of length $N=200$ bases with equal probability for
each of the four nucleotides.  The resulting protein-protein correlation
function, shown in Fig.~\ref{fig:logCor_Vienna}, again follows a power law.

We would like to point out that, even though our simplified model and
the Vienna package render very similar exponents for the power-law
correlation, we do not give much credence to the precise value of
this value, since it is dependend on strong finite-size effects.
E.g., it appears that the effective exponent depends somewhat on
protein concentrations and its absolute value decreases 
(i.e. the power law decays slower) as the ratios $c_k/K_{d,k}^{(0)}$
increase. The reason for such tendency is still unclear and requires
more investigation in the future.  However, independent of the precise
value of the exponent, the important fact is that the protein-protein
correlation function does not decay exponentially but rather has a fat
(power-law) tail which enables interdependency between
proteins binding at long distances from each other along the molecule.

\section{Conclusion}

In summary, we have proposed a mechanism for long-range interactions between multiple binding partners of an RNA molecule. According to our model, the ensemble of RNA secondary structures can be viewed as a medium for this long range interaction. When one protein or microRNA binds to the RNA, it also changes the ensemble, and the change of the ensemble transmits the effect of this binding partner to the other binding partner, thus resulting in an interplay between them. Following this concept, we have quantified this interplay at least for the simplest two-partner case. We considered coarse-grained models for RNA secondary structures, and discovered that a long-range power-law protein-protein correlation function occurs when a loop penalty is taken into account in the model. 

We discussed this discovery in detail using analytical and numerical approaches.  For the RNA in the high-temperature molten phase, we analytically derived a power-law correlation function, and verified this result numerically. In the low-temperature glass phase, we numerically calculated the linear correlation function and discovered that this long-range phenomenon is strongly affected by protein concentrations, but occurs precisely at those biologically meaningful protein concentrations at which the RNA is neither fully saturated with proteins nor completely unoccupied. In Biology, this interdependency between protein binding sites thus can play an important role in combinatorial gene regulation on the post-transcriptional level. It is therefore worth to theoretically and experimentally investigate the phenomenon discovered here on the level of generic sequences for specific naturally occurring RNA molecules and their protein binding partners.

The conclusions of the coarse-grained models are verified by the Vienna package, which is viewed as the state of the art for the numerical simulation of real RNA molecules. Some of the advanced discussion about the loop penalty, leading to the issue of its length dependence, has been performed, and a similar power-law linear correlation function was shown. We thus believe that the constant loop initiation penalty, the one we focused on in the very first discussion, is the critical factor for the occurrence of the long-range effect. 

In our discussions we assumed that the proteins do not directly
interact with each other in order to focus on the role of the RNA
secondary structure alone.  Based on our calculations, the structure
mediated indeterdependcy between binding sites is due to
configurations in which the protein
binding sites are located on opposite sides of the same stem of the
molecule.  Thus, in these configurations the proteins are in fact in
close spatial proximity and direct interactions between the proteins
would further enhance the interdependency observed here.

In consequence, our work is an initial discovery and discussion of
this RNA-mediated interaction. Still many issues, such as the exact
glass-phase exponent of the power law, the effect of additional
details in RNA folding models, and establishing that this RNA
structure mediated interdependency between binding sites 
is in fact used in post-transcriptional regulation of real mRNAs, are
unsolved and left for future investigations.


\section{Acknowledgements}

We acknowledge fruitful discussions with Nikolaus Rajewsky that initiated this work. This material is based upon work supported by the National Science Foundation under Grant No. DMR-01105458.


\appendix

\section{Molten-phase partition function of the model including a constant loop cost}
	\label{app:ZN}

In the molten phase, an RNA molecule can be viewed as a homopolymer, and the partition functions $Z_{(i,j)}$ and $Z_{b(i,j)}$ in Eq. (\ref{eq:recursion}) retard to functions of sequence length $j-i+1$.  Its partition function has been calculated before~\cite{Muller03,Liu04,Tamm07} but we include it here for the sake of completeness.

Due to the translational invariance, we can rewrite the partition functions as
\begin{equation}
Z_b(j-i+1) = Z_{b(i,j)} \; , \; \mathrm{and} \; Z_0(j-i+1) = Z_{(i,j)}.
\end{equation}
The recursive relation in Eq. (\ref{eq:recursion}) is then rewritten as
\begin{subequations}
\begin{align}
Z_b(N\!+\!1) & = q Z_b(N\!-\!1) + q \xi \left( Z_0(N\!-\!1) - Z_b(N\!-\!1) \right), \label{eq:ini_recursive1} \\ 
Z_0(N\!+\!1) & = Z_0(N) + \sum_{k=0}^{N\!-\!1} Z_0(k) Z_b(N\!-\!k\!+\!1) \label{eq:ini_recursive2}.
\end{align}
\end{subequations}
The first of these equations can be solved for $G$ as
\begin{equation}
Z_0(N) = \frac{1}{q \xi} Z_b(N+2) - \frac{1-\xi}{\xi} Z_b(N),
\end{equation}
and can then be used to replace all $Z_0$'s in the second equation, rendering a purely $Z_b$-dependent recursion. Applying z-transformation to this recursion with the initial conditions
\begin{equation}
Z_b(0) = 0, \;  Z_b(1) = 0, \;  Z_b(2) = q \xi, 
\end{equation}
yields the equation for the z-space partition function $\hat{Z}_b(z) \equiv \sum_{N=0}^{\infty} Z_b(N) z^{-N}$ 
\begin{equation}
\left(\frac{z^2}{q\xi}\!-\!\frac{1\!-\!\xi}{\xi} \right) \hat{Z}_b(z)^2 - \frac{1}{z}\left(\frac{z^2}{q\xi} \!-\! \frac{1\!-\!\xi}{\xi} \right) (z-1) \hat{Z}_b(z) + 1 \!=\! 0.
\end{equation}
The limited partition function in z-space is thus given as
\begin{equation}
\hat{Z}_b(z) = \frac{z-1}{2z} - \frac{1}{2} \sqrt{\frac{(z-1)^2}{z^2} - \frac{4q\xi}{(z^2 - q(1-\xi))}}. 
	\label{eq:Bz_complete}
\end{equation}
The z-space partition function $\hat{Z}_0(z)$ can be derived by applying z-transformation to Eq. (\ref{eq:ini_recursive1}), yielding
\begin{equation} 
\hat{Z}_0(z) = \left( \frac{z^2}{q\xi}  - \frac{1-\xi}{\xi} \right) \hat{Z}_b(z),
	\label{eq:Gz-Bz}
\end{equation}
which leads to $\hat{Z}_0(z)$ by substituting Eq. (\ref{eq:Bz_complete}) for $\hat{Z}_b(z)$, yielding
\begin{equation}
\begin{aligned}
\hat{Z}_0(z)  & = \frac{(z-1)(z^2 - q + q\xi)}{2 q \xi z} \\
& - \frac{\sqrt{ \left[ z^2 - q+q\xi \right]\left[ (z-1)^2 ( z^2 - q+q\xi)  - 4 q \xi z^2  \right]  }}{2 q\xi z}.
\end{aligned}
	\label{eq:Gz_final}
\end{equation}
The partition function in real space is then derived by the inverse z-transformation,
\begin{equation}
Z_0(N) = \frac{1}{2\pi i} \oint_C dz z^{N-1} \hat{Z}_0(z).
\end{equation}
As $N \rightarrow \infty $, the leading order term of this contour integral, contributed by the topological defect with the largest real part of $z$, determines its value. According to Eq. (\ref{eq:Gz_final}), $\hat{Z}_0(z)$ has a simple pole at $z_p = 0$, and branch cuts at all $z$ satisfying
\begin{equation}
\left[ z^2 - q+q\xi \right]\left[ (z-1)^2 ( z^2 - q+q\xi)  - 4 q \xi z^2  \right] \leq 0.
	\label{eq:eqBC}
\end{equation}
The first component in Eq. (\ref{eq:eqBC}) leads to two branch cut end points $z_{\pm} = \pm \sqrt{q(1-\xi)}$. Substituting these two values into the function of the second component,
\begin{equation}
f(z) = (z-1)^2 ( z^2 - q+q\xi)  - 4 q \xi z^2 
\end{equation}
it can be found that $f(z_{\pm}) < 0$. Since $f(\infty) > 0$ is known, at least one of the branch cuts must extend along the real axis beyond $z = z_+$. Following the derivation in the appendix of Ref. \cite{Liu04}, the large-$N$ expression of $G(N)$ is then obtained as
\begin{equation} 
Z_0(N) = A(q,\xi) N^{-3/2} z_c^N(q, \xi)[1 + O(N^{-1})], 
	\label{eq:G_largeN}
\end{equation}
where $z_c$ is the solution of $f(z_c) = 0$ with the greatest real part, and the prefactor $A(q,\xi)$ is given by
\begin{equation}
A = \frac{\sqrt{z_c(z_c^2-q+q\xi) f'(z_c) }}{2\pi z_c q \xi} \Gamma\left( \frac{3}{2} \right). 
\end{equation}
To simplify the notation, we omit the subscript $c$ in this manuscript
and write down the asymptotic partition function,
for example in Eq.~(\ref{eq:G_BC}), using $z$ rather than $z_c$.


\section{Calculation of the molten-phase correlation function for the model including a constant loop cost to first order in loop cost}
	\label{app:gD_calculation}

In this appendix, we will calculate the correlation function $g(D)$
between two protein binding sites at a distance $D$ from each other.
To calculate the correlation function $g(D)$, we need to know the
changed components $\Zdinstem$, $\Zddtwostem$, and $\Zddinstem{1}{1}$
in the limited partition functions $Z_d$ and $Z_{dd}$.  Since these
are difficult to obtain exactly, we will only calculate their
expansions in $(1-\xi)$, thus taking the effects of a finite loop cost
into account perturbatively.  In addition, to exclude all boundary and
finite-size effects, we consider the limit of an infinitely long
molecule in which both footprints are far from the ends of the RNA,
i.e., the limit of $N \gtrsim n_1 \approx n_2 \gg D \gg l \geq 1$.
Here, we will aim to calculate the overall correlation function $g(D)$
to first order in $(1-\xi)$ which will show its power law dependence on
the distance $D$ between the binding sites.
Appendix~\ref{app:second_order} will then demonstrate that this power
law dependence is not changed by the second order term, which is a lot
more difficult to obtain.

The limited partition functions $\Zdinstem$ and $\Zddtwostem$
appear in Eqs.~(\ref{eq:Zd-S-Z0}) and~(\ref{eq:Zdd_final}) with
prefactors of $(1-\xi)$ while $\Zddinstem{1}{1}$ occurs
with a prefactor of $(1-\xi)^2$.  Thus, an expansion of the correlation
function $g(D)$ to first order in $(1-\xi)$ requires knowledge
of $\Zdinstem$ and $\Zddtwostem$ to zeroth order in $(1-\xi)$ but
does not depend on $\Zddinstem{1}{1}$.  We will thus start by calculating
the limited partition functions $\Zdinstem$ and $\Zddtwostem$ to
zeroth order in $(1-\xi)$.

\subsection{$\Zdinstem$, the partition function for the changed structures in $Z_d$}\label{app:Zdinstem_0th}

\begin{figure}
$\begin{array}{c}
\subfloat[]{\includegraphics[width=0.95\columnwidth]{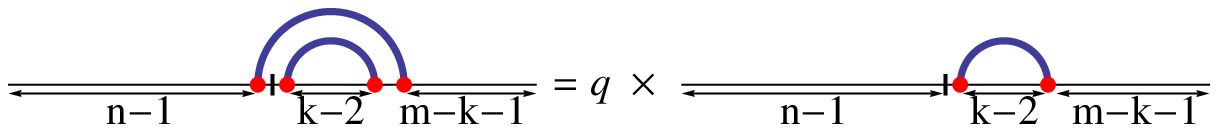} \label{fig:Cdstrategy1} } \\
\subfloat[]{\includegraphics[width=0.95\columnwidth]{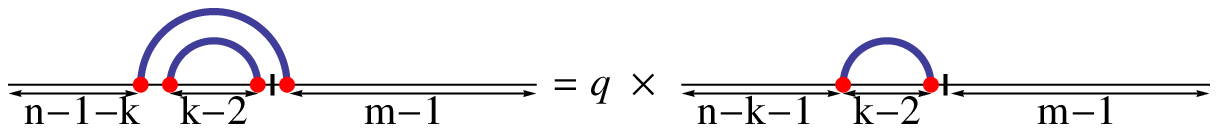} \label{fig:Cdstrategy2} }
\end{array}$
\caption{A one-to-one mapping from the changed structures to the structures including the certain bond. The footprint is between the $n^{\mathrm{th}}$ and the $(n+1)^{\mathrm{th}}$. Red dots are the nucleotides forming the base pair stack which includes the footprint. The two types of structures are taken into account in Eq. (\ref{eq:Zdinstem_initial}), where the first summation is for the structures in (a) and the second one is for (b). All labels for the segment lengths in the plot do not count the red dots.}
	\label{fig:Cdstrategy}
\end{figure}

The partition function for the changed structures in $Z_d(n,m)$,
expressed as $\Zdinstem(n,m)$, includes the structures shown in
Fig. \ref{fig:Zdstem}, i.e., those structures in $Z_0(n+m)$ in which
the $n^{th}$ and the $(n+1)^{st}$ base are involved in a base stack of
a stem. To calculate the contribution of all these structures we note
that in all structures in which the $n^{th}$ and the $(n+1)^{st}$ base
are involved in a base stack the two consecutive base pairs can be
contracted to a single base pair yielding a new structure on an RNA of
$n+m-2$ bases, in which the stem is shortened by one stack but the
topology of the structure is otherwise unchanged.  This process is
described in Fig.~\ref{fig:Cdstrategy}. Thus, qualitatively
$\Zdinstem(n,m)$ is given as $q$ times (to represent the one additional
base pair) the partition function of all structures of an $n+m-2$ base
RNA in which the $n^{th}$ base is paired; this in turn is the
partition function over all structures of an $n+m-2$ base RNA minus the
partition function over all structures of an $n+m-2$ base RNA in which
the $n^{th}$ base is unpaired and the latter is in turn the partition
function over all structures of an $n+m-3$ base RNA, i.e., we would
expect
\begin{displaymath}
\Zdinstem(n,m) \approx q \left[ Z_0(n+m-2) - Z_0(n+m-3) \right].
\end{displaymath}

However, there are some subleties missed in the above qualitative
argument that require a more careful computation.  To this end, we
write $\Zdinstem(n,m)$ explicitly as a sum over the position of the
other side of the stack that the $n^{th}$ and $(n+1)^{st}$ base are
included in and separate this sum into terms where the other side of
the stack is to the left of $n$ and terms where the other side of the
stack is to the right of $n$ as shown in Fig.~\ref{fig:Cdstrategy}.
This yields
\begin{equation}
\begin{aligned}
\Zdinstem(n,m) & = q \sum_{k=2}^{n-1} Z_b(k) \ZtwosegV{n-k-1}{m-1} \\
& + q \sum_{k=2}^{m-1} Z_b(k) \ZtwosegV{n-1}{m-k-1}, 
\end{aligned}
	\label{eq:Zdinstem_initial}
\end{equation}
where $\ZtwosegV{k_1}{k_2}$ is for the outer part in
Fig. \ref{fig:Cdstrategy}, taking into account all structures formed
by the nucleotides in the two \textit{non-independent} segments of
lengths $k_1$ and $k_2$. Notice that $\Ztwoseg$ is symmetric
with respect to its two segments, i.e. $\ZtwosegV{k_1}{k_2} =
\ZtwosegV{k_2}{k_1}$.

For zero loop cost ($\xi=1$) we would simply have
$\ZtwosegV{k_1}{k_2}=Z_0(k_1+k_2)$.  Since our goal is here
to calculate only the zeroth-order term of $\Zdinstem$, we can thus
use
\begin{equation}
\ZtwosegV{k_1}{k_2} = Z_0(k_1+k_2) + O((1-\xi)).
	\label{eq:Ztwoseg_simple}
\end{equation}
Similarly, Eq.~(\ref{eq:ini_recursive1}) yields
\begin{equation}
\begin{aligned}
Z_b(k) & = q\xi Z_0(k-2) + q(1-\xi) Z_b(k-2)
	\label{eq:Zb-to-Z0} \\
& = q\xi Z_0(k-2) + O((1-\xi)).
\end{aligned}
\end{equation}

Substituting the zeroth-order approximated $\Ztwoseg$ and $Z_b$ into
Eq.~(\ref{eq:Zdinstem_initial}), $\Zdinstem$ can be approximated to the
zeroth order in $(1-\xi)$ as
\begin{equation}
\begin{aligned}
& \Zdinstem(n,m) \\
& = q^2 \xi \left( \sum_{k=2}^{m-1} +  \sum_{k=2}^{n-1} \right) Z_0(k-2) Z_0(n+m-k-2) \\
& \quad + O((1-\xi)) \\
& = q^2 \xi \sum_{k=2}^{n-1} Z_0(k-2) Z_0(n+m-k-2) \\
& \quad + q^2 \xi \sum_{k' = n+1}^{n+m-2} Z_0(n+m-k'-2) Z_0(k'-2) \\
& \quad + O((1-\xi)) \\
& = q \sum_{k=2}^{n+m-2} \left[ q \xi Z_0(k-2) \right] Z_0(n+m-2-k) \\
& \qquad  - q^2 \xi Z_0(n-2) Z_0(m-2) + O((1-\xi)) \\
& = q \left[ \sum_{k=2}^{n+m-2} Z_b(k)Z_0(n+m-2-k) \right] \\
& \qquad  - q^2 \xi Z_0(n-2) Z_0(m-2) + O((1-\xi)) .
\end{aligned}
	\label{eq:Zdinstem_simplified}.
\end{equation}
The summation term is exactly equivalent to a conditional partition
function of an RNA molecule with $n+m-2$ nucleotides which takes into
account all structures in which the $1^{\mathrm{st}}$ nucleotide is paired
with another arbitrary nucleotide. This conditional partition function
can be calculated by subtracting the partition function for
the structures including an unpaired $1^{\mathrm{st}}$ nucleotide from
the whole partition function, leading to
\begin{equation}
 \sum_{k=2}^{n\!+\!m\!-\!2} Z_b(k)Z_0(n+m-2-k) \!=\! Z_0(n+m-2)\!-\!Z_0(n+m-3).
 	\label{eq:cyclicRNA_view}
\end{equation}
Thus, the partition function for the changed structures, $\Zdinstem$,
is derived to the zeroth order in $(1-\xi)$ as
\begin{equation}
\begin{aligned}
\Zdinstem(n,m) = & q \left[ Z_0(n+m-2) - Z_0(n+m-3) \right] \\
& - q^2 \xi Z_0(n-2)Z_0(m-2)  \\
& + O((1-\xi)),
\end{aligned}
	\label{eq:Cd_with_Z0_final}
\end{equation}
which is nearly our naive expectation.  In fact, in the relevant
limit of $n\approx m\approx N/2\to\infty$, we can insert the asymptotic form
Eq.~(\ref{eq:G_BC}) of $Z_0$ to obtain
\begin{equation}
\begin{aligned}
\Zdinstem(n,m) = & q A \left[ \frac{z^{n+m-2}}{(n+m-2)^{3/2}} -
\frac{z^{n+m-3}}{(n+m-3)^{3/2}} \right] \\
& -q^2 \xi A^2 \frac{z^{n+m-4}}{(n-2)^{3/2}(m-2)^{3/2}} \\
& + O((1-\xi),z^{n+m}N^{-5/2})
\end{aligned}
\end{equation}
and find that the additional term on the second line is actually
decaying with a power of $N^{-3}$ compared to the power of
$N^{-{3/2}}$ of the terms on the first line, which represent our naive
expectation, and thus can be neglected (the $z^{n+m}$ behavior in the
numerator is the same for all terms).  This finally yields
\begin{equation}
\begin{aligned}
\Zdinstem(n,m) = & q A z^{n+m}\frac{z-1}{z^3}N^{-3/2} \\
& + O((1-\xi),z^{n+m}N^{-5/2})
\end{aligned}
\label{eq:Cd_final}
\end{equation}


\subsection{$\Zddtwostem$, contributions to $Z_{dd}$ when both footprints
are in the same base stack}\label{app:Zddtwostem_0th}

\begin{figure}
\includegraphics[width=0.7\columnwidth]{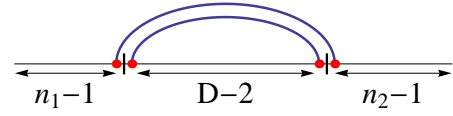} 
\caption{The structures included in $\Zddtwostem$. The outer segment
  lengths $n_1-1$ and $n_2-1$ do not include the two nucleotides
  forming the outer bond. The partition function for the inner segment
  is given by $Z_b(D)$ and thus includes an algebraic
  component $D^{-3/2}$ for $D \gg 1$. }
	\label{fig:Zddtwostem} 
\end{figure}

The structures included in $\Zddtwostem$, in which both footprints are
in the same stack of a stem, are shown in Fig. \ref{fig:Zddtwostem}.
Upon contracting the two stacked base pairs into one, these
configurations exactly correspond to the configurations of an RNA of
$n_1+D+n_2-2$ bases in which base $n_1$ and base $n_1+D-2$ are
paired.  Thus, this partition function encodes the pairing probability
for two bases of an RNA with distance $D-2$ which is known to depend
like a power law on the distance $D$.

Specifically, the partition function for the structures included in this
quantity can be written as
\begin{equation}  
\Zddtwostem(n_1,D,n_2) = q Z_b(D) \ZtwosegV{n_1-1}{n_2-1}.
	\label{eq:Zddtwostem_def}
\end{equation}

Since we only need to know the zeroth order term of $\Zddtwostem$
in $(1-\xi)$ we can substitute the zeroth order expansions
Eqs.~(\ref{eq:Ztwoseg_simple}) and~(\ref{eq:Zb-to-Z0}) of $\Ztwoseg$
and $Z_b$ respectively and obtain
\begin{equation}
\begin{aligned}
 \Zddtwostem = & q^2 \xi Z_0(D-2) Z_0(n_1+n_2-2) \\ 
& + O((1-\xi)).
	\label{eq:Zddtwostem_0th-approx}
\end{aligned}
\end{equation}
Inserting the asymptotic form Eq.~(\ref{eq:G_BC})
for $Z_0$ finally yields
\begin{equation}
\begin{aligned}
& \Zddtwostem(n_1,D,n_2) \\
& = \frac{q^2 \xi }{z^4}
A^2\frac{z^{n_1+n_2+D}}{D^{3/2}N^{3/2}}\\
& \quad + O((1-\xi),z^{n_1+D+n_2}N^{-5/2},z^{n_1+D+n_2}D^{-5/2}).
\end{aligned}
\label{eq:Zddtwostem_0th_asympt}
\end{equation}
This term explicitly contains the power law depence on the distance $D$
between the protein binding sites.


\subsection{Correlation function}

The molten-phase protein-protein correlation function, $g(D)$, is
given by the first equality in Eq. (\ref{eq:g12-molten}) in terms of
the limited partition functions $Z_d$ and $Z_{dd}$, and the
protein-binding parameters $c_i$ and $K_i$ with $i = 1,2$. As we
have described at the beginning of this appendix, to exclude
all boundary and finite-size effects, we consider the limit of an
infinitely long molecule in which both footprints are far from the
ends of the RNA, i.e., the limit of $N \gtrsim n_1 \approx n_2 \gg D
\gg l \geq 1$.

As a first step to calculating $g(D)$
we divide the numerator and the denominator of the first expression
in Eq.~(\ref{eq:g12-molten}) by $Z_0^2$ and get
\begin{equation}
\begin{aligned}
g(D) & = \frac{\frac{Z_0 Z_{dd} - Z_d \times Z_d}{Z_0^2}}{\frac{Z^2}{Z_0^2}} \\
& = \frac{\frac{Z_{dd}}{Z_0} - \frac{Z_d}{Z_0} \times \frac{Z_d}{Z_0} }{\left( 1 + \frac{c_1}{K_{d,1}^{(0)}} \frac{Z_d}{Z_0} +\frac{c_2}{K_{d,2}^{(0)}} \frac{Z_d}{Z_0} + \frac{c_1 c_2}{K_{d,1}^{(0)}K_{d,1}^{(0)}} \frac{Z_{dd}}{Z_0} \right)^2}\\
&\equiv \frac{{\cal N}}{{\cal D}^2}.
\end{aligned}
	\label{eq:gD_fraction}
\end{equation}
Thus, the relevant quantities are the ratios $Z_d/Z_0$ and
$Z_{dd}/Z_0$, which are obtained by dividing the results from
sections~\ref{app:Zdinstem_0th} and~\ref{app:Zddtwostem_0th} by the
asymptotic form $Z_0(N)\approx A z^N/N^{3/2}$. It is easy to see that
this division eliminates the exponential dependence on $z$ and cancels
the $N^{-3/2}$ dependence of the results from
sections~\ref{app:Zdinstem_0th} and~\ref{app:Zddtwostem_0th}.

We will first calculate the numerator and denominator of $g(D)$
separately to appropriate orders in $(1-\xi)$, and then merge them
together to derive $g(D)$ to the first order in $(1-\xi)$.  We are
going to show that $g(D)$ decays algebraically as $D \gg 1$ and thus
supports a long-range interaction between binding proteins. 
Appendix~\ref{app:second_order} will then demonstrate that this
algebraic behavior also holds up in the second order in $(1-\xi)$.

The numerator ${\cal N}$ of $g(D)$ is the difference between two
products of two partition functions. Without loop cost, this
difference, as shown in Eq. (\ref{eq:gD_simplest_zero}), leads to a
residue in $O(1/N^2)$ and thus converges to zero as $N \to \infty$.
In the model including a constant loop cost, however, the residue is
$O(1/N^0)$ and does not diminish in the large-$N$ limit. The goal in
this section is to calculate this finite residue, and confirm that
this residue is a power-law function of $D$ and therefore supports a
long-range effect in the system.

With the help of the relations in Eqs.~(\ref{eq:Zd-S-Z0})
and~(\ref{eq:Zdd_final}), the numerator in Eq. (\ref{eq:gD_fraction}) 
is written as
\begin{equation}
\begin{aligned}
{\cal N} & = \frac{Z_{dd}(n_1,D,n_2)}{Z_0(N)} \\
& \qquad - \frac{Z_d(n_1,D+l+n_2)}{Z_0(N)} \frac{Z_d(n_1+l+D,n_2)}{Z_0(N)} \\
= &  \left[ \frac{Z_0(N-2l)}{Z_0(N)} - \left(\frac{Z_0(N-l)}{Z_0(N)}\right)^2 \right] \\
& - (1-\xi) \times \\
& \quad \left[ \frac{\Zdinstem(n_1,D+n_2)}{Z_0(N)} - \frac{Z_0(N-l)}{Z_0(N)} \frac{\Zdinstem(n_1,D+l+n_2)}{Z_0(N)} \right] \\
& - (1-\xi) \times \\
& \quad \left[ \frac{\Zdinstem(n_1+D, n_2)}{Z_0(N)} - \frac{Z_0(N-l)}{Z_0(N)} \frac{\Zdinstem(n_1+D+l, n_2)}{Z_0(N)} \right] \\
& + (1-\xi)^2 \times \\
& \quad \left[ \frac{ \Zddinstem{1}{1}}{Z_0(N)} - \frac{\Zdinstem(n_1,D+l+n_2)}{Z_0(N)}\frac{\Zdinstem(n_1+l+D,n_2)}{Z_0(N)} \right] \\
&  + \xi (1-\xi) \frac{ \Zddtwostem}{Z_0(N)} .
\end{aligned}
	\label{eq:gD_numerator}
\end{equation} 
The first term is the same as for the model in the absence of a loop
cost and thus leads to a residue proportional to $(l/N)^2$ which
vanishes in the limit $N \to \infty$.  The forth term is of second
order in $(1-\xi)$ and can thus be ignored here.  To calculate the
second term, we substitute the asymptotic expressions
Eqs.~(\ref{eq:G_BC}) and~(\ref{eq:Cd_final}) for $Z_0$ and
$\Zdinstem$, respectively, and obtain for the term in brackets
\begin{equation}
\begin{aligned}
&\left[ \frac{\Zdinstem(n_1,D+n_2)}{Z_0(N)} - \frac{Z_0(N-l)}{Z_0(N)} \frac{\Zdinstem(n_1,D+l+n_2)}{Z_0(N)} \right] \\
=&
q\frac{z-1}{z^{3+2l}}\left[\frac{N^{3/2}}{(N-2l)^{3/2}}-\frac{N^3}{(N-l)^3}\right]+O(N^{-1})
\end{aligned}
\end{equation}
which also vanishes for $N\to\infty$.  Due to symmetry the same argument
applies to the third term in Eq.~(\ref{eq:gD_numerator}).

Finally, we can substitute the asymptotic expansions
Eqs.~(\ref{eq:G_BC}) and~(\ref{eq:Zddtwostem_0th_asympt}) of $Z_0$ and
$\Zddtwostem$, respectively, into the last term to find the entire numerator
as
\begin{equation}
{\cal N} =  (1-\xi) \frac{q^2\xi^2A}{z^{4+2l}}D^{-3/2}+O((1-\xi)^2,D^{-5/2},N^{-1}).
	\label{eq:gD_num_large_1st_simplified}
\end{equation}

Considering that this numerator of $g(D)$ has an explicit prefactor of
$(1-\xi)$, it is enough to calculate the denominator to zeroth order
in $(1-\xi)$.  Including all the arguments of the limited partition
functions, the denominator is given by (before taking the square)
\begin{equation}
\begin{aligned}
{\cal D}=&  1 \!+\! \frac{c_1}{K_{d,1}^{(0)}} \frac{Z_d(n_1,D\!+\! l \!+\!n_2 )}{Z_0(N)} \\
& \!+\! \frac{c_2}{K_{d,2}^{(0)}} \frac{Z_d(n_1\!+\! l \!+\!D,n_2 )}{Z_0(N)} \!+\! \frac{c_1 c_2}{K_{d,1}^{(0)}K_{d,2}^{(0)}} \frac{Z_{dd}(n_1,D,n_2)}{Z_0(N)},
\end{aligned}
	\label{eq:g_down_initial}
\end{equation}
Eqs.~(\ref{eq:Zd-S-Z0}) and~(\ref{eq:Zdd_final}) explicitly show that
in zeroth order in $(1-\xi)$ $Z_d$ and $Z_{dd}$ can be replaced by
$Z_0$ thus yielding
\begin{equation}
\begin{aligned}
{\cal D}=&  1 + \frac{c_1}{K_{d,1}^{(0)}} \frac{Z_0(N-l)}{Z_0(N)}
 + \frac{c_2}{K_{d,2}^{(0)}} \frac{Z_0(N-l)}{Z_0(N)}\\
&\quad + \frac{c_1 c_2}{K_{d,1}^{(0)}K_{d,2}^{(0)}} \frac{Z_0(N-2l)}{Z_0(N)}
+O((1-\xi))\\
=&  1 + \frac{c_1}{K_{d,1}^{(0)}} \frac{N^{3/2}}{(N-l)^{3/2}z^l}
 + \frac{c_2}{K_{d,2}^{(0)}} \frac{N^{3/2}}{(N-l)^{3/2}z^l}\\
&\quad + \frac{c_1 c_2}{K_{d,1}^{(0)}K_{d,2}^{(0)}} \frac{N^{3/2}}{(N-2l)^{3/2}z^{2l}}
+O((1-\xi),N^{-1})\\
& = \left(\!1 + \frac{c_1}{K_{d,1}^{(0)}z^l} \!\right)
\left(\! 1 + \frac{c_2}{K_{d,2}^{(0)}z^l} \!\right) +O((1-\xi),N^{-1}).
\end{aligned}
	\label{eq:gD_den_large_1st_simplified}
\end{equation}
where we have again used the asymptotic expression Eq.~(\ref{eq:G_BC})
for $Z_0$.

The correlation function is then obtained by dividing
Eq.~(\ref{eq:gD_num_large_1st_simplified}) by the square of
Eq.~(\ref{eq:gD_den_large_1st_simplified}) and is thus given by
\begin{equation}
\begin{aligned}
g(D)=&(1-\xi)\frac{q^2\xi^2A}{z^{4+2l}\left(1 + \frac{c_1}{K_{d,1}^{(0)}z^l}\right)^2
\left( 1 + \frac{c_2}{K_{d,2}^{(0)}z^l}\right)^2}D^{-3/2}\\
&\quad + O((1-\xi),N^{-1},D^{-5/2}).
\end{aligned}
\end{equation}

Consequently, once a finite loop cost is added into the RNA-folding model, 
a long-range correlation occurs between two binding partners on the RNA.


\section{Calculation of the molten-phase correlation function for the model including a constant loop cost to second order in loop cost} \label{app:second_order}

In this appendix, we will calculate the second order terms in the
expansion of the correlation function $g(D)$ in $(1-\xi)$, i.e., in
the loop energy.  Since the calculation of this term is quite
involved, it is important to point out that the main result, the power
law behavior of the correlation function $g(D)$, already occurs in the
first order in $(1-\xi)$ as detailed in
Appendix~\ref{app:gD_calculation}. We do include the calculation here
nevertheless for two reasons. First, it quantitatively improves the
pre-factor of the power law when compared to numerical results at
finite $(1-\xi)$.  Second, the fact that the second order term has the
same power law dependence on the distance $D$ as the first order term
strengthens the argument that this power law behavior is not simply an
artifact of the perturbative calculation.  However, the reader content
with only the expansion of the correlation function to first order in
$(1-\xi)$ may opt to skip this appendix.


\subsection{$\Zdinstem$, the partition function for the changed structures in $Z_d$}\label{app:Zdinstem_1st}

Calculating the correlation function $g(D)$ to second order requires
expanding the limited partition function $\Zdinstem$ to first order
in $(1-\xi)$.  Our starting point for this calculation will be
Eq.~(\ref{eq:Zdinstem_initial}).  To make progress, we need to
know the partition function $\Ztwoseg$ to first order in $(1-\xi)$.

\begin{figure}
$\begin{array}{c}
\subfloat[]{\includegraphics[width=0.95\columnwidth]{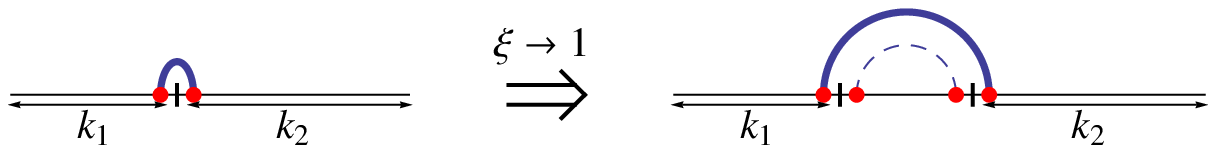} \label{fig:Ztwoseg1} } \\
\subfloat[]{\includegraphics[width=0.95\columnwidth]{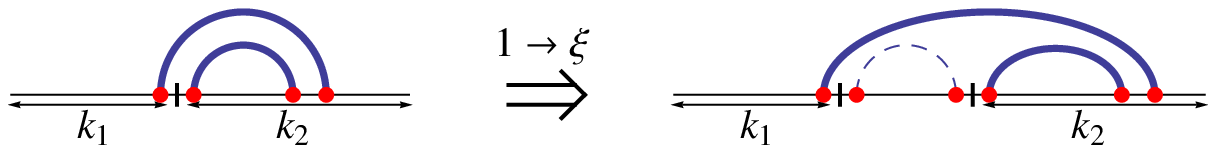} \label{fig:Ztwoseg2} }
\end{array}$
\caption{Two types of structures which contribute differently in $Z_0(k_1+k_2)$ (left) and $\ZtwosegV{k_1}{k_2}$ (right). Solids lines are base pairs in $Z_0(k_1,k_2)$ and $\ZtwosegV{k_1}{k_2}$, and dashed lines are the bonds in the right hand side of Fig. \ref{fig:Cdstrategy}, whose effects are necessary to be considered in the calculation of $\ZtwosegV{k_1}{k_2}$. (a) A hairpin loop in $Z_0(k_1 + k_2)$ becomes a base pair stack in a stem in $\ZtwosegV{k_1}{k_2}$. (b) A stack in a stem in $Z_0(k_1 + k_2)$ becomes part of a multiloop in $\ZtwosegV{k_1}{k_2}$. }
	\label{fig:Ztwoseg}
\end{figure}

To find the expansion of $\ZtwosegV{k_1}{k_2}$, again
two groups of secondary structures have to be distinguished in
$\ZtwosegV{k_1}{k_2}$.  Similar to the idea of calculating
$\Zdinstem(n,m)$, one group of structures contributes the same in both
$Z_0(k_1+k_2)$ and $\ZtwosegV{k_1}{k_2}$, and the other contributes
differently.  The latter group includes two types of structures, whose
differences in contribution between $Z_0(k_1+k_2)$ and
$\ZtwosegV{k_1}{k_2}$ are shown in Fig.~\ref{fig:Ztwoseg}.  Thus,
$\ZtwosegV{k_1}{k_2}$ can be expressed as $Z_0(k_1+k_2)$ plus a
changed term resulting from the loop cost as
\begin{equation}
\begin{aligned}
\ZtwosegV{k_1}{k_2} & = Z_0(k_1+k_2) \\
& + (1-\xi) \left[ q \ZtwosegV{k_1-1}{k_2-1} - \Zdinstem(k_1,k_2) \right],
\end{aligned}
	\label{eq:Ztwoseg_complete}
\end{equation}
where the term including $\ZtwosegV{k_1-1}{k_2-1}$ contains
the configurations in Fig. \ref{fig:Ztwoseg}(a) and
the following term is for those in Fig. \ref{fig:Ztwoseg}(b).
Substituting Eq.~(\ref{eq:Ztwoseg_complete})  into Eq.~(\ref{eq:Zdinstem_initial}) rewrites $\Zdinstem$ as
\begin{subequations}
\begin{align}
&\Zdinstem(n,m) \nonumber \\
=& q \left( \sum_{k=2}^{n-1} + \sum_{k=2}^{m-1}\right) Z_b(k) Z_0(n+m-k-2) \label{eq:Zdinstem1st_1} \\
& + q^2(1-\xi) \sum_{k=2}^{n-2} Z_b(k)\ZtwosegV{n-k-2}{m-2} \label{eq:Zdinstem1st_2} \\
& + q^2(1-\xi) \sum_{k=2}^{m-2} Z_b(k)\ZtwosegV{n-2}{m-k-2} \label{eq:Zdinstem1st_3} \\
& - q(1-\xi) \sum_{k=2}^{n-2} Z_b(k) \Zdinstem(n-k-1, m-1) \label{eq:Zdinstem1st_4} \\
& - q(1-\xi) \sum_{k=2}^{m-2} Z_b(k) \Zdinstem(m-k-1, n-1) \label{eq:Zdinstem1st_5}.
\end{align}
	\label{eq:Zdinstem1st}
\end{subequations}

We will now calculate each term in Eq.~(\ref{eq:Zdinstem1st}) to the
first order in $(1-\xi)$. Note, that all summations in these terms are
multiplied by $(1-\xi)$, except the first
term~(\ref{eq:Zdinstem1st_1}). Therefore, our calculation will be to
the first order for the summation in (\ref{eq:Zdinstem1st_1}), and to
the zeroth order for the remaining ones, i.e.,
terms~(\ref{eq:Zdinstem1st_2})-(\ref{eq:Zdinstem1st_5}).

We first notice that the combination of terms~(\ref{eq:Zdinstem1st_2}) and~(\ref{eq:Zdinstem1st_3}) has the same form as the expression of $\Zdinstem$ in Eq.~(\ref{eq:Zdinstem_initial}), and thus can be written as 
$q (1-\xi) \Zdinstem(n-1,m-1)$. 
Substituting the zeroth-order $\Zdinstem$ from Eq.~(\ref{eq:Cd_with_Z0_final}) leads to the first-order expression for the combination of the two terms,
\begin{equation}
\begin{aligned}
& q^2(1-\xi) \sum_{k=2}^{n-2} Z_b(k)\ZtwosegV{n-k-2}{m-2}\\
& + q^2(1-\xi) \sum_{k=2}^{m-2} Z_b(k)\ZtwosegV{n-2}{m-k-2}\\
=& q^2(1-\xi)[Z_0(n+m-4) - Z_0(n+m-5)] \\
& - q^3 \xi (1-\xi)Z_0(n-3) Z_0(m-3) \\
& + O((1-\xi)^2). \\
\end{aligned}
	\label{eq:Zdinstem1st_23_final}
\end{equation}


Next, we calculate the first one among the remaining three terms,
i.e., term~(\ref{eq:Zdinstem1st_1}). To evaluate this term to the first
order in $(1-\xi)$, it is necessary to first figure out the partition
function $Z_b$ to first order. Iterating Eq.~(\ref{eq:Zb-to-Z0}) once
leads to the approximation,
\begin{equation}
Z_b(D) = q\xi Z_0(D-2) + q^2\xi(1-\xi) Z_0(D-4) + O((1-\xi)^2 ).
\label{eq:Zb-firstorder}
\end{equation}
Substituting this approximation into the second summation in
term~(\ref{eq:Zdinstem1st_1}) yields the first order approximation
\begin{equation}
\begin{aligned}
& q \sum_{k=2}^{m-1} Z_b(k) Z_0(n+m-k-2) \\ 
= & q^2 \xi \sum_{k=2}^{m-1} Z_0(k-2) Z_0(n+m-k-2) \\
   & + q^3 \xi (1-\xi)\sum_{k=4}^{m-1} Z_0(k-4) Z_0(n+m-k-2) \\
   & + O((1-\xi)^2). \\
\end{aligned}
	\label{eq:Zdinstem1st_1_sum-a2}
\end{equation}
We first calculate the first term in Eq.~(\ref{eq:Zdinstem1st_1_sum-a2}). Applying the changing variable strategy similar to that in Eq.~(\ref{eq:Zdinstem_simplified}), this term becomes
\begin{equation}
\begin{aligned}
& q^2 \xi \sum_{k=2}^{m-1} Z_0(k-2) Z_0(n+m-k-2) \\
= & q^2 \xi \sum_{k' = n+1}^{n+m-2} Z_0(n+m-k'-2) Z_0(k'-2) \\
= & q \sum_{k=n+1}^{n+m-2} Z_b(k) Z_0(n+m-k-2) \\
   & - q^3 \xi (1-\xi) \sum_{k=n+1}^{n+m-2} Z_0(k-4) Z_0(n+m-k-2) \\
   & + O((1-\xi)^2) \\
= & q \sum_{k=n+1}^{n+m-2} Z_b(k) Z_0(n+m-k-2) \\
   & - q^3 \xi (1-\xi) \sum_{k=4}^{m+1} Z_0(k-4) Z_0(n+m-k-2) \\
   & + O((1-\xi)^2) 
\end{aligned}
	\label{eq:Zdinstem1st_1_sum-a2-step}
\end{equation}
Substituting Eq.~(\ref{eq:Zdinstem1st_1_sum-a2-step}) into Eq.~(\ref{eq:Zdinstem1st_1_sum-a2}) results in a subtraction between two first-order summations and leads to
\begin{equation}
\begin{aligned}
& q \sum_{k=2}^{m-1} Z_b(k) Z_0(n+m-k-2) \\ 
= & q \sum_{k=n+1}^{n\!+\!m\!-\!2} Z_b(k) Z_0(n+m-k-2) \\
   & + q^3 \xi (1-\xi)\left( \sum_{k=4}^{m-1} - \sum_{k=4}^{m+1}\right) Z_0(k\!-\!4) Z_0(n\!+\!m\!-\!k\!-\!2) \\
   & + O((1-\xi)^2)\\
= & q \sum_{k=n+1}^{n\!+\!m\!-\!2} Z_b(k) Z_0(n\!+\!m\!-\!k\!-\!2) \\
   & - q^3 \xi(1\!-\!\xi) [Z_0(m\!-\!4) Z_0(n\!-\!2) + Z_0(m\!-\!3)Z_0(n\!-\!3)] \\
   & + O((1-\xi)^2) 
\end{aligned}
	\label{eq:Zdinstem1st_1_sum-a2-simplified}
\end{equation}
Finally, substituting Eq.~(\ref{eq:Zdinstem1st_1_sum-a2-simplified}) into term~(\ref{eq:Zdinstem1st_1}) yields
\begin{equation}
\begin{aligned}
& q \left( \sum_{k=2}^{n-1} + \sum_{k=2}^{m-1}\right) Z_b(k) Z_0(n+m-k-2)\\
= & q \left( \sum_{k=2}^{n-1} + \sum_{k=n+1}^{n+m-2} \right) Z_b(k) Z_0(n+m-k-2) \\
   & - q^3 \xi(1\!-\!\xi) [Z_0(m\!-\!4) Z_0(n\!-\!2) \!+\! Z_0(m\!-\!3)Z_0(n\!-\!3)] \\
   & + O((1-\xi)^2) \\
= & q \sum_{k=2}^{n+m-2} Z_b(k) Z_0(n+m-k-2) \\
   & - q [ q \xi Z_0(n-2) + q^2 \xi (1-\xi) Z_0(n-4) ] Z_0(m-2) \\
   & - q^3 \xi(1\!-\!\xi) [Z_0(m\!-\!4) Z_0(n\!-\!2) \!+\! Z_0(m\!-\!3)Z_0(n\!-\!3)] \\ 
   & + O((1-\xi)^2). \\
\end{aligned}
	\label{eq:Zdinstem1st_1_rewritten}
\end{equation}
With the help of the equality in Eq.~(\ref{eq:cyclicRNA_view}), Eq.~(\ref{eq:Zdinstem1st_1_rewritten}) can be further simplified to
\begin{equation}
\begin{aligned}
& q \left( \sum_{k=2}^{n-1} + \sum_{k=2}^{m-1}\right) Z_b(k) Z_0(n+m-k-2)\\
=& q[ Z_0(n+m-2)-Z_0(n+m-3)] \\ 
& - q^2 \xi Z_0(n-2) Z_0(m-2) \\
& - q^3 \xi (1-\xi) [ Z_0(n-4) Z_0(m-2) + \\
& \qquad \qquad\quad Z_0(n-3)Z_0(m-3) \!+\! Z_0(n-2)Z_0(m-4) ] \\
& + O((1-\xi)^2). \\
\end{aligned}
	\label{eq:Zdinstem1st_1_final}
\end{equation}

There are now two last terms, terms~(\ref{eq:Zdinstem1st_4})
and~(\ref{eq:Zdinstem1st_5}), remaining to be calculated. To this end,
we substitute the zeroth-order $\Zdinstem$ in
Eq. (\ref{eq:Cd_with_Z0_final}), yielding
\begin{equation}
\begin{aligned}
& - q(1-\xi) \sum_{k=2}^{n-2} Z_b(k) \Zdinstem(n-k-1, m-1)\\
& - q(1-\xi) \sum_{k=2}^{m-2} Z_b(k) \Zdinstem(m-k-1, n-1)\\
= & -q^2(1-\xi)\left( \sum_{k=2}^{n-2} + \sum_{k=2}^{m-2}\right) Z_b(k) Z_0(n+m-k-4) \\
& +q^2(1-\xi)\left( \sum_{k=2}^{n-2} + \sum_{k=2}^{m-2}\right) Z_b(k) Z_0(n+m-k-5) \\
& + q^3 \xi(1-\xi) \sum_{k=2}^{n-3} Z_b(k) Z_0(n-k-3)Z_0(m-3) \\
& + q^3 \xi(1-\xi) \sum_{k=2}^{m-3} Z_b(k) Z_0(m-k-3)Z_0(n-3) \\
& + O((1-\xi)^2).
\end{aligned}
	\label{eq:Zdinstem1st_45_step1}
\end{equation}
The first two terms in Eq.~(\ref{eq:Zdinstem1st_45_step1}) are both in
the form of the zeroth-order $\Zdinstem$ in
Eq.~(\ref{eq:Zdinstem_simplified}) (considering $Z_b(k) = q\xi
Z_0(k-2) + O((1-\xi))$) and thus their combination can be rewritten as
\begin{equation}
\begin{aligned}
& \!-\! q (1-\xi) \Zdinstem(n\!-\!1,m\!-\!1) \\
& \!+\! q (1-\xi) \Zdinstem(n\!-\!2,m\!-\!1) \!+\! q^2 (1\!-\!\xi) Z_b(n\!-\!2) Z_0(m\!-\!3) \\
& \!+\! O((1-\xi)^2) \\
= & \!-\! q^2 (1-\xi)[Z_0(n+m-4) \\
& \qquad \qquad -2Z_0(n+m-5) + Z_0(n+m-6) ] \\
   & \!+\! q^3 \xi (1-\xi) Z_0(n-3) Z_0(m-3) \\
   & \!+\! O((1-\xi)^2) 
\end{aligned}
	\label{eq:Zdinstem1st_45_step1-1}
\end{equation}
The last two terms can be simplified by rewriting the summations with the help of the relation in Eq.~(\ref{eq:cyclicRNA_view}), which yields
\begin{equation}
\begin{aligned}
& q^3 \xi (1-\xi) [ Z_0(n-3) - Z_0(n-4)] Z_0(m-3) \\
& + q^3 \xi (1-\xi) [ Z_0(m-3) - Z_0(m-4)] Z_0(n-3) \\
= & q^3 \xi (1-\xi)[  2Z_0(n-3)Z_0(m-3) \\
& \qquad\qquad + Z_0(n\!-\!3)Z_0(m\!-\!4) \!+\! Z_0(n\!-\!4)Z_0(m\!-\!3) ].
\end{aligned}
	\label{eq:Zdinstem1st_45_step1-2}
\end{equation}
Collecting Eqs.~(\ref{eq:Zdinstem1st_45_step1-1})
and~(\ref{eq:Zdinstem1st_45_step1-2}), we express the combination of
terms~(\ref{eq:Zdinstem1st_4}) and~(\ref{eq:Zdinstem1st_5}) to the
first order in $(1-\xi)$ as
\begin{equation}
\begin{aligned}
& - q(1-\xi) \sum_{k=2}^{n-2} Z_b(k) \Zdinstem(n-k-1, m-1)\\
& - q(1-\xi) \sum_{k=2}^{m-2} Z_b(k) \Zdinstem(m-k-1, n-1)\\
= & -q^2 (1-\xi)[Z_0(n+m-4) \\
& \qquad \qquad -2Z_0(n+m-5) + Z_0(n+m-6) ] \\
& + q^3 \xi (1-\xi)[ 3Z_0(n-3) Z_0(m-3) \\
& \qquad\qquad \!+\! Z_0(n-3)Z_0(m-4) \!+\! Z_0(n-4)Z_0(m-3) ] \\
& + O((1-\xi)^2) 
\end{aligned}
	\label{eq:Zdinstem1st_45_final}
\end{equation}
All five terms in Eq.~(\ref{eq:Zdinstem1st}) are now calculated to the
first order in $(1-\xi)$. We then collect these terms in
Eqs.~(\ref{eq:Zdinstem1st_23_final}),~(\ref{eq:Zdinstem1st_1_final}),
and~(\ref{eq:Zdinstem1st_45_final}) and discover the first order
expression of $\Zdinstem(n,m)$ as
\begin{widetext}
\begin{equation}
\begin{aligned}
\Zdinstem(n,m) & = q[Z_0(n+m-2)-Z_0(n+m-3)] - q^2 \xi Z_0(n-2)Z_0(m-2) \\
& +  q^2(1-\xi)[Z_0(n+m-5)-Z_0(n+m-6)] \\
& - q^3 \xi (1-\xi) [Z_0(n-4)Z_0(m-2) + Z_0(n-2) Z_0(m-4) - Z_0(n-3)Z_0(m-3) ] \\
& + q^3 \xi (1-\xi) [Z_0(n-3)Z_0(m-4) + Z_0(n-4)Z_0(m-3)] \\
& + O((1-\xi)^2), 
\end{aligned}
	\label{eq:Zdinstem1st_final_Zs}
\end{equation}
\end{widetext}
where the first line is the zeroth-order term, exactly identical to the one derived in Eq.~(\ref{eq:Cd_with_Z0_final}), and all the subsequent terms are the first order term in $(1-\xi)$.

In the limit $N/2 \approx n \approx m \gg 1$, the asymptotic form of Eq.~(\ref{eq:Zdinstem1st_final_Zs}) is given by inserting $Z_0(N) \approx A z^N N^{-3/2}$.
As before, all terms in which $n$ and $m$ are not arguments of the same $Z_0$
decay as $N^{-3}$ and can thus be neglected with respect to the $N^{-3/2}$
dependence of the terms where $n+m$ is the argument of one $Z_0$.
The asymptotic $\Zdinstem$ is thus given by the remaining relevant terms as
\begin{equation}
\begin{aligned}
\Zdinstem(n,m) = & q A z^{n+m} \frac{z-1}{z^3} N^{-3/2} \\
& + q^2 A (1-\xi) z^{n+m} \frac{z-1}{z^6} N^{-3/2} \\
& + O((1-\xi)^2, z^{n+m}N^{-5/2}).
\end{aligned}
	\label{eq:Zdinstem1st_final}
\end{equation}



\subsection{$\Zddtwostem$, contributions to $Z_{dd}$ when both footprints
are in the same base stack}

The starting point for the first order terms of the limited partition
function $\Zddtwostem$ is Eq.~(\ref{eq:Zddtwostem_def}). We obtain the
expansion to first order in $(1-\xi)$ by substituting the first order
expansions of $Z_b$ and $\Ztwoseg$.  The first order expansion of $Z_b$ has
already been given in Eq.~(\ref{eq:Zb-firstorder}) and we obtain the
first order expression of $\Ztwoseg$ by substituting the zeroth-order
$\Zdinstem$ from Eq.~(\ref{eq:Cd_with_Z0_final}) into
Eq.~(\ref{eq:Ztwoseg_complete}) yielding
\begin{equation}
\begin{aligned}
\ZtwosegV{k_1}{k_2} & = Z_0(k_1+k_2) \\
& + (1-\xi) \left[ q Z_0(k_1+k_2-3)  \right. \\
& \qquad \qquad + \left. q^2\xi Z_0(k_1-2)Z_0(k_2-2) \right] \\
& + O((1-\xi)^2).
\end{aligned}
\label{eq:Ztwoseg_1st}
\end{equation}
This substitution reveals the first order expansion of the limited partition
function $\Zddtwostem$ as
\begin{equation}
\begin{aligned}
 \Zddtwostem = & q^2 \xi Z_0(D-2) Z_0(n_1+n_2-2) \\ 
& + q^3 \xi (1-\xi) Z_0(D-4) Z_0(n_1+n_2-2) \\
& + q^3 \xi (1-\xi) Z_0(D-2) Z_0(n_1+n_2-5) \\
& + q^4\xi^2 (1-\xi) Z_0(D-2) Z_0(n_1-3)Z_0(n_2-3) \\
& + O((1-\xi)^2).
	\label{eq:Zddtwostem_1st-approx}
\end{aligned}
\end{equation}
In the limit $n_1\approx n_2\approx N/2\to\infty$,
the last term can be dropped since it is of higher order in $1/N$
than the others.  Inserting the asymptotic form Eq.~(\ref{eq:G_BC})
for $Z_0$ finally yields
\begin{equation}
\begin{aligned}
& \Zddtwostem(n_1,D,n_2) \\
& = \left( \frac{q^2 \xi }{z^4} + (1-\xi)\frac{q^3 \xi(z-1)}{z^7} \right)
A^2\frac{z^{n_1+n_2+D}}{D^{3/2}N^{3/2}}\\
& \quad + O((1-\xi)^2,z^{n_1+D+n_2}N^{-5/2},z^{n_1+D+n_2}D^{-5/2}).
\end{aligned}
\label{eq:Zddtwostem_first_final}
\end{equation}


\subsection{$\Zddinstem{1}{1}$, contributions to $Z_{dd}$ when both footprints
are in stacks}

The limited partition function $\Zddinstem{1}{1}$ over all
configurations in which both footprints are inserted into base stacks
is multiplied in the expression for $Z_{dd}$ by $(1-\xi)^2$.
Therefore, it was not relevant when calculating the correlation
function $g(D)$ to first order in $(1-\xi)$ but needs to be
considered to zeroth order in $(1-\xi)$, now that we aim for the
second order term of the correlation function $g(D)$.

\begin{figure}
$\begin{array}{c}
\subfloat[]{\includegraphics[width=0.95\columnwidth]{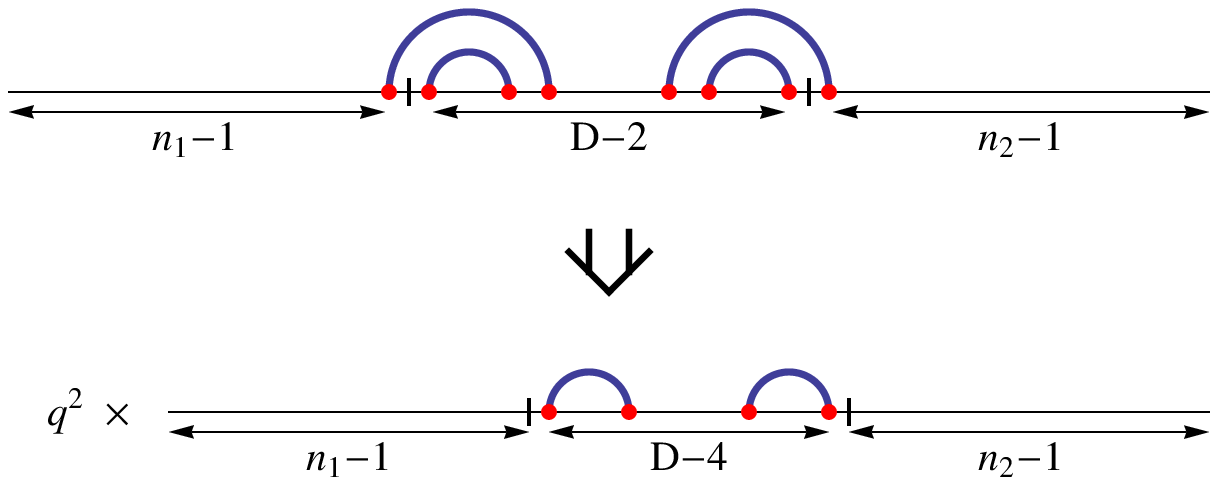} \label{fig:Zddinstem11naive1} } \\
\subfloat[]{\includegraphics[width=0.95\columnwidth]{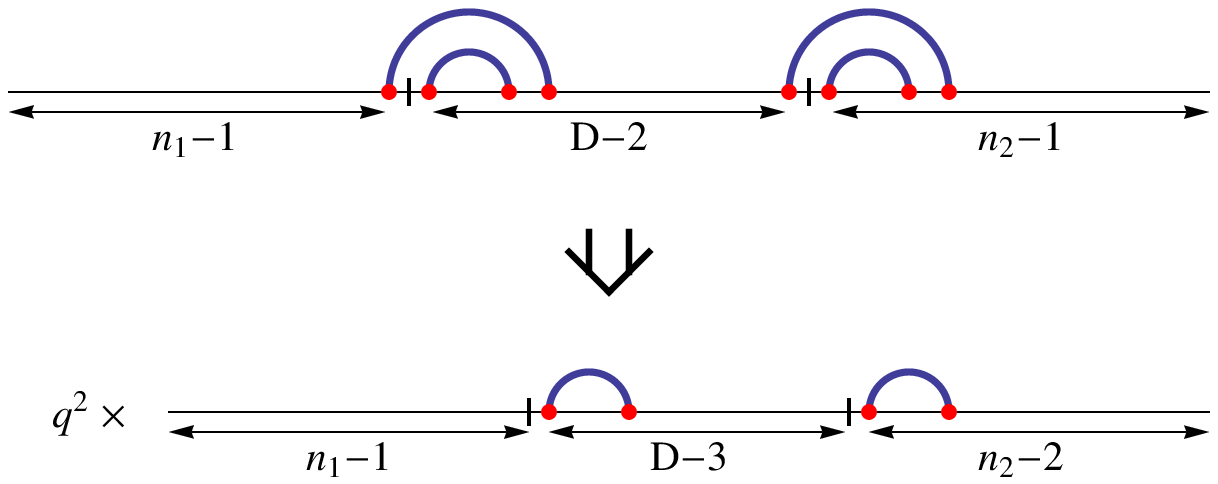} \label{fig:Zddinstem11naive2} }
\end{array}$
\caption{Two examples for the naive expectation for $\Zddinstem{1}{1}$. In these two examples, the contractions from base pair stacks to single base pairs result in the same topology, whereas the explicit positions of the remaining base pairs and the distances between two inserted footprints are different. These small deviations from the $n_1^{\mathrm{th}}$ and $(n_1+D)^{\mathrm{th}}$ base pairs are ignored in the naive estimation for $\Zddinstem{1}{1}$. }
	\label{fig:Zddinstem11naive}
\end{figure}

Qualitatively, $\Zddinstem{1}{1}$ can be estimated by a strategy
similar to the one that yielded the naive expectation for
$\Zdinstem$. That is, $\Zddinstem{1}{1}$ can be roughly given as $q^2$
times the partition function for all structures of an $n_1+D+n_2-4$
base RNA in which the $n_1^{\mathrm{th}}$ and $(n_1+D)^{\mathrm{th}}$
bases are required to be paired (albeit not necessarily with each
other), as described in Fig. \ref{fig:Zddinstem11naive} for two
examples.  The latter structures can in turn be calculated by starting
from the partition function over all structures of an $n_1+D+n_2-4$
base RNA, subtracting all those in which the $n_1^{th}$ or the
$(n_1+D)^{th}$ base are unpaired and adding back the structures that
were subtracted twice because both bases are unpaired.  Thus, we would
expect
\begin{equation}
\begin{aligned}
\Zddinstem{1}{1} & \approx q^2 [ Z_0(n_1+D+n_2-4) - 2Z_0(n_1+D+n_2-5) \\
& \qquad + Z_0(n_1+D+n_2-6)].
	\label{eq:Zddinstem11_naive_ini}
\end{aligned}
\end{equation}

However, this naive estimation again has shortcomings.  It is
not always exactly the $n_1^{\mathrm{th}}$ and $(n_1+D)^{\mathrm{th}}$
base which are paired with other bases and not even their distance is
always exactly $D$ bases.  Based on the configurations of the original
structure with two inserted base pair stacks, the exact position of
the two paired bases (in the structures where the inserted stacks are
contracted to single bonds) can deviate from the $n_1^{\mathrm{th}}$
and $(n_1+D)^{\mathrm{th}}$ by up to $\pm 4$.  E.g.,
Fig.~\ref{fig:Zddinstem11naive} describes two configurations in which
the two paired bases are at different positions and with different
deviations.  Just as in the case of $\Zdinstem(n,m)$ (see
Eq.~(\ref{eq:Cd_with_Z0_final})) this leads to additional terms.  However,
while in the case of $\Zdinstem(n,m)$ these additional terms became
irrelevant in the limit $n_1\approx n_2\approx N/2\to\infty$, here
some of the terms remain relevant and contain $Z_0(D)$, thus
contributing to the power law dependence on $D$.

\begin{figure}
$\begin{array}{c}
\subfloat[]{\includegraphics[width=0.7\columnwidth]{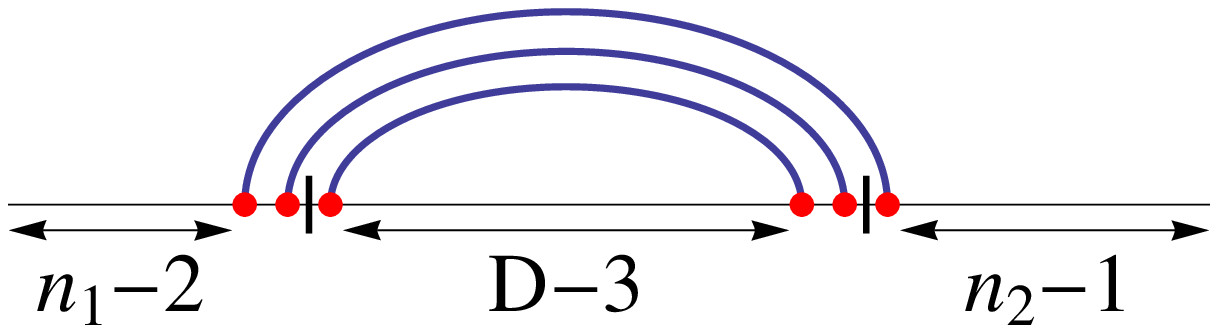} \label{fig:Zddinstem11_4} } \\
\subfloat[]{\includegraphics[width=0.7\columnwidth]{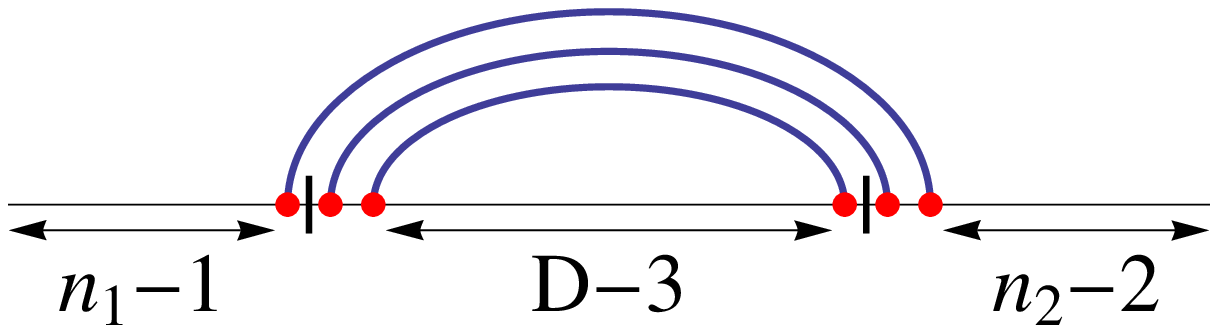} \label{fig:Zddinstem11_5} }  
\end{array}$
\caption{Two configurations included in $\Zddinstem{1}{1}$. The two
  footprints are in two consecutive stacks of the same stem, sharing a
  mutual bond. The partition function of these two configurations is
  very similar to $\Zddtwostem$, in which the two footprints are in
  the same stack.}
	\label{fig:Zddinstem11mutual}
\end{figure}

\begin{figure}
$\begin{array}{c}
\subfloat[]{\includegraphics[width=0.95\columnwidth]{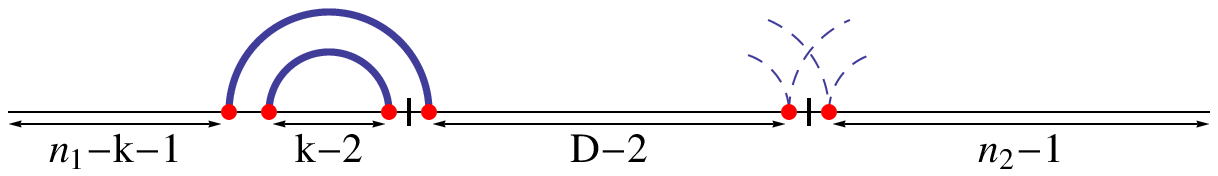} \label{fig:Zddinstem11_1} } \\
\subfloat[]{\includegraphics[width=0.95\columnwidth]{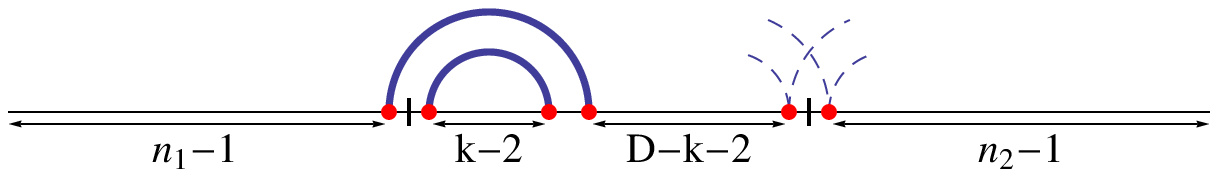} \label{fig:Zddinstem11_2} } \\
\subfloat[]{\includegraphics[width=0.95\columnwidth]{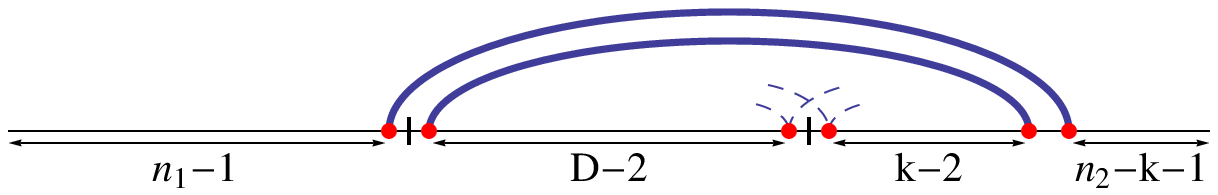} \label{fig:Zddinstem11_3} } 
\end{array}$
\caption{The three types of structures included in
  $\Zddinstem{1}{1}-\Zddtwostem-\ZddinstemMutual$, i.e., when both
  footprints are in different stems or different base pair stacks of
  the same stem. Two footprints are inserted between the
  $n_1^{\mathrm{th}}$ and $(n_1+1)^{\mathrm{st}}$ and the
  $(n_1+D)^{\mathrm{th}}$ and $(n_1+D+1)^{\mathrm{st}}$
  nucleotide. All labels for segment lengths in the figures do not
  take into account the red dots, which form the bonds nearby the
  footprints.}
	\label{fig:Zddinstem11}
\end{figure}

To explicitly derive the partition function $\Zddinstem{1}{1}$ 
to the zeroth order in $(1-\xi)$,
we separate the structures included in $\Zddinstem{1}{1}$
into three types of configurations, described in
Figs. \ref{fig:Zddtwostem}, \ref{fig:Zddinstem11mutual}, and
\ref{fig:Zddinstem11}, respectively. In all secondary structures
described in these three figures, both footprints are in either
different or the same base pair stack(s). However, they share the
bonds of the base pair stack(s) in different ways.

In Fig. \ref{fig:Zddtwostem}, both footprints are in the same base
stack of a stem. The partition function for this configuration is
exactly $\Zddtwostem$, the zeroth order expansion of which has
already been given in Eq.~(\ref{eq:Zddtwostem_0th-approx}).

In Fig. \ref{fig:Zddinstem11mutual}, the two footprints are in two
consecutive stacks of the same stem, and thus share a mutual
bond. Considering the similarity of Figs. \ref{fig:Zddinstem11_4} and
\ref{fig:Zddinstem11_5} with \ref{fig:Zddtwostem}, their partition
functions should be very similar to $\Zddtwostem$.  In fact, the
partition function for the structures in Fig. \ref{fig:Zddinstem11_4}
is given as
\begin{equation}
\begin{aligned}
& q^2 Z_b(D-1) \ZtwosegV{n_1-2}{n_2-1} \\
= & q^3 \xi Z_0(D-3) Z_0(n_1+n_2-3) + O((1-\xi)),
\end{aligned}
\end{equation}
where the zeroth-order approximations $\ZtwosegV{n_1-2}{n_2-1}=
Z_0(n_1+n_2-3)+O((1-\xi))$ (Eq.~(\ref{eq:Ztwoseg_simple})) and
$Z_b(D-1)= q \xi Z_0(D-3)+O((1-\xi))$ (Eq.~(\ref{eq:Zb-to-Z0})) have
been applied. The partition function for the structures in
Fig. \ref{fig:Zddinstem11_5} is obtained by exchanging the
variables $n_1$ and $n_2$ which leads to the same result to zeroth
order in $(1-\xi)$. Thus, the partition function for all structures
including a mutual bond, which are described in
Fig. \ref{fig:Zddinstem11mutual}, is given as
\begin{equation}
\begin{aligned}
\ZddinstemMutual = & 2q^3 \xi Z_0(D-3) Z_0(n_1+n_2-3) \\
& + O((1-\xi))
\end{aligned}
\label{eq:ZddinstemMutual}
\end{equation}
to the zeroth order in $(1-\xi)$

The remaining structures in $\Zddinstem{1}{1}$ are described in
Fig. \ref{fig:Zddinstem11}. The three Figs. \ref{fig:Zddinstem11_1},
\ref{fig:Zddinstem11_2}, and \ref{fig:Zddinstem11_3} enumerate all
structures in $\Zddinstem{1}{1}-\Zddtwostem-\ZddinstemMutual$ by
the following strategy: all three possible configurations
of the base pair stack comprising the first (left) footprint are
shown separately in the three parts of the figure, and in each of these
configurations all possible configurations of the second (right)
footprint are considered. Notice that since $\Zddinstem{1}{1}$ has to
be a symmetric function of $n_1$ and $n_2$, evaluating either the left
or the right footprint as the ``first" footprint makes no difference. 
We then define the partition functions for the structures considered in
Figs. \ref{fig:Zddinstem11_1}, \ref{fig:Zddinstem11_2}, and
\ref{fig:Zddinstem11_3} as $\ZddinstemOne$, $ \ZddinstemTwo$, and $
\ZddinstemThree$ respectively, and calculate the partition function
for the remaining structures as 
$\Zddinstem{1}{1}-\Zddtwostem-\ZddinstemMutual =
\ZddinstemOne+ \ZddinstemTwo+ \ZddinstemThree$.

\begin{figure}
$\begin{array}{c}
\subfloat[]{\includegraphics[width=0.95\columnwidth]{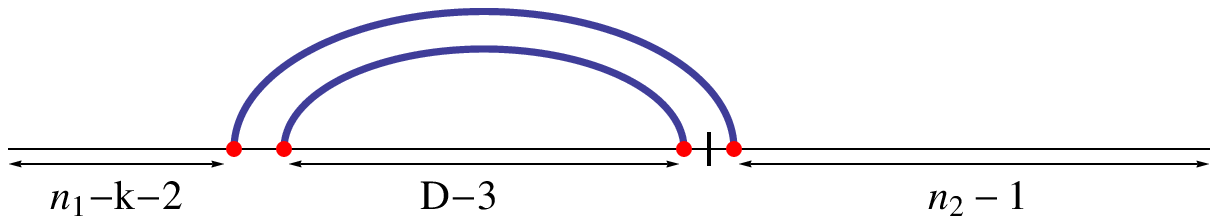} \label{fig:Zddinstem11of1Fix1} } \\
\subfloat[]{\includegraphics[width=0.95\columnwidth]{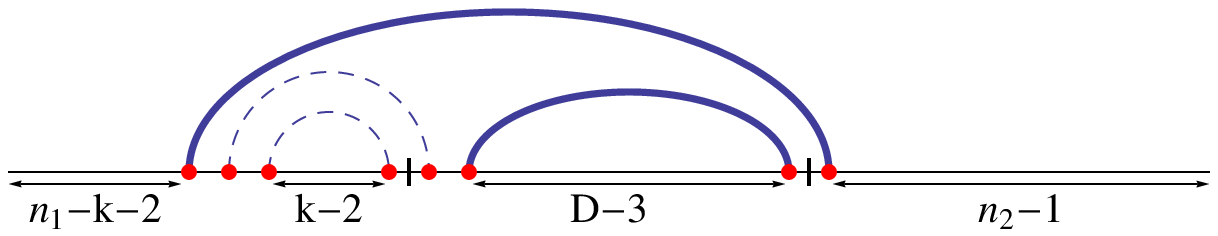} \label{fig:Zddinstem11of1Fix2} } \\
\end{array}$
\caption{The structures which (a) contribute differently in $Z_0$ and $Z_d$ and thus included in $\Zdinstem$ but (b) contribute identically in the corresponding $\Ztwoseg$ and $\Zdtwoseg$ and thus should not be taken into account in $\Ztwoseginstem$. These structure have to be excluded when approximating $\Ztwoseginstem$ as $\Zdinstem$.}
	\label{fig:Zddinstem11of1Fix}
\end{figure}

The partition function for the structures included in Fig. \ref{fig:Zddinstem11_1} 
is written down as
\begin{equation}
\ZddinstemOne = q \sum_{k=2}^{n-1} Z_b(k) \ZtwoseginstemV{n_1-k-1}{D-1}{n_2},
	\label{eq:h1_inital}
\end{equation}
where $\ZtwoseginstemV{n_1-k-1}{D-1}{n_2}$ is the partition function
for all structures formed by the nucleotides outside of the base pair
stack containing the first footprint, with the condition that the
$(n_1+D)^{\mathrm{th}}$ and $(n_1+D+1)^{\mathrm{st}}$ nucleotide are
one side of a base pair stack.  We would like to calculate
$\ZtwoseginstemV{n_1-k-1}{D-1}{n_2}$ by simply removing the inserted
stack and enclosed base pairs after the $(n_1-k-1)^{st}$ base since
this would yield the quantity $\Zdinstem(n_1+D-k-2,n_2)$ which we have
calculated before.  However, there are two effects of removing the
inserted stack.  First, the removal will lead to replacements of a
factor $\xi$ by $1$ (or vice versa) in the loop or stack containing
the removed section if loops are turned into stacks or vice versa.
Since we are only interested in the zeroth order in $(1-\xi)$ this can
be ignored.  Second, and more importantly, after removal of the left
stack, the structures shown in Fig.~\ref{fig:Zddinstem11of1Fix}(b), in
which the right insertion site is not part of a base stack and which
are thus not part of $\Zddinstem{1}{1}$ turn into the structures shown
in Fig.~\ref{fig:Zddinstem11of1Fix}(a) which thus {\em are} included
in $\Zdinstem(n_1-k+D-2,n_2)$.  Thus, their contribution
$qZ_b(D-1)\ZtwosegV{n_1-k-2}{n_2-1}$ needs to be subtracted yielding
to zeroth order in $(1-\xi)$
\begin{equation}
\begin{aligned}
& \ZtwoseginstemV{n_1-k-1}{D-1}{n_2} \\
& = \Zdinstem(n_1-k+D-2,n_2) \\
& \quad - q^2\xi Z_0(D-3)Z_0(n_1+n_2-k-3)+ O((1-\xi))\\
& = q[Z_0(n_1\!+\!n_2\!+\!D-k-4)-Z_0(n_1\!+\!n_2\!+\!D-k-5)]\\
& \quad -q^2\xi Z_0(n_1+D-k-4)Z_0(n_2-2)\\
& \quad - q^2\xi Z_0(D-3)Z_0(n_1+n_2-k-3)+ O((1-\xi))
\end{aligned}
	\label{eq:Ztwoseginstem_approx}
\end{equation}
where we have used Eq.~(\ref{eq:Cd_with_Z0_final}) in the second equality.  Inserting
this into Eq.~(\ref{eq:h1_inital}) and using Eq.~(\ref{eq:Zb-to-Z0})
for the zeroth order approximation of $Z_b(k)$ we get
 
\begin{subequations}
\begin{align}
\ZddinstemOne = & q^3 \xi \sum_{k=2}^{n_1\!-\!1} Z_0(k-2) Z_0(n_1\!+\!n_2\!+\!D\!-\!k\!-\!4)  \label{eq:ZddinstemOne1} \\
& - q^3 \xi \sum_{k=2}^{n_1\!-\!1} Z_0(k\!-\!2) Z_0(n_1\!+\!n_2\!+\!D\!-\!k\!-\!5)
	\label{eq:ZddinstemOne2} \\
& - q^4 \xi^2 Z_0(n_2\!-\!2) \sum_{k=2}^{n_1\!-\!1} Z_0(k\!-\!2) Z_0(n_1\!+\!D\!-\!k\!-\!4) 
	\label{eq:ZddinstemOne3} \\
& - q^4 \xi^2  Z_0(D\!-\!3) \sum_{k=2}^{n_1\!-\!2} Z_0(k\!-\!2)Z_0(n_1\!+\!n_2\!-\!k\!-\!3) 
	\label{eq:ZddinstemOne4} \\
& + O((1-\xi)) \notag
\end{align}
	\label{eq:ZddinstemOne}
\end{subequations}
Notice that the last summation is up to $k = n_1-2$ instead of $k = n_1-1$
as for the other terms since the subtraction of the terms shown in
Fig.~\ref{fig:Zddinstem11of1Fix} is not necessary in the case $k=n_1-1$.

Similarly, the partition function for the structures in Fig.~\ref{fig:Zddinstem11_2} is written down as
\begin{equation}
\ZddinstemTwo = q \sum_{k=2}^{D-2} Z_b(k) \ZtwoseginstemV{n_1-1}{D-k-1}{n_2}.
	\label{eq:h2_inital}
\end{equation}
Again, the two factors can be replaced by their zeroth order terms
using Eqs.~(\ref{eq:Zb-to-Z0}) and~(\ref{eq:Ztwoseginstem_approx})
yielding
\begin{subequations}
\begin{align}
\ZddinstemTwo = & q^3 \xi \sum_{k=2}^{D-2} Z_0(k\!-\!2) Z_0(n_1\!+\!n_2\!+\!D\!-\!k\!-\!4) 
	\label{eq:ZddinstemTwo1} \\
& -  q^3 \xi \sum_{k=2}^{D-2} Z_0(k\!-\!2) Z_0(n_1\!+\!n_2\!+\!D\!-\!k\!-\!5)  
	\label{eq:ZddinstemTwo2} \\
& - q^4 \xi^2 Z_0(n_2\!-\!2) \sum_{k=2}^{D-2} Z_0(k\!-\!2) Z_0(n_1\!+\!D\!-\!k\!-\!4) 
	\label{eq:ZddinstemTwo3} \\
& - q^4 \xi^2 Z_0(n_1\!+\!n_2\!-\!3) \sum_{k=2}^{D-3} Z_0(k\!-\!2) Z_0(D\!-\!k\!-\!3) 
	\label{eq:ZddinstemTwo4} \\
& + O((1-\xi)). \notag
\end{align}
	\label{eq:ZddinstemTwo}
\end{subequations}

The partition function for the structures described in
Fig.~\ref{fig:Zddinstem11_3}
can be written down as
\begin{equation}
\ZddinstemThree = q^2 \sum_{k=2}^{n_2-1} \Zdinstemenclosed(D-1, k-1) \ZtwosegV{n_1-1}{n_2-k-1},
	\label{eq:h3_inital}
\end{equation}
where $\Zdinstemenclosed$ is a partition function over the same structures as
in $\Zdinstem$ with the only difference that the weights of structures in
$\Zdinstemenclosed$ are calculated in the context of an enclosing base
pair (the one from base $n_1+1$ to base $n_1+D+k$ in
Fig.~\ref{fig:Zddinstem11_3}) while the weights of the structures
in $\Zdinstem$ are evaluated in an open context.  The context of the
enclosing base pair implies that the weights of nearly all structures
get multiplied by $\xi$ for the outermost loop closed by that enclosing
base pair with the exception of the structures in which the first and
last base of the substrand described by $\Zdinstemenclosed$ are paired.
The latter structures in turn contain an $\Zdinstemenclosed$ on a shortened
sequence, i.e.
\begin{equation}
\Zdinstemenclosed(n,m)=\xi\Zdinstem(n,m)+(1-\xi)q\Zdinstemenclosed(n-1,m-1).
\end{equation}
To the zeroth order in $(1-\xi)$ we may neglect  the second term and
thus find
\begin{subequations}
\begin{align}
\ZddinstemThree = & q^2 \xi\sum_{k=2}^{n_2\!-\!1} \Zdinstem(D\!-\!1, k\!-\!1) \ZtwosegV{n_1\!-\!1}{n_2\!-\!k\!-\!1} \\
& +O((1-\xi))\notag\\
= & q^3 \xi  \sum_{k=2}^{n_2\!-\!1} Z_0(n_1\!+\!n_2\!-\!k\!-\!2) Z_0(D\!+\!k\!-\!4)
\label{eq:ZddinstemThree1} \\
& - q^3 \xi  \sum_{k=2}^{n_2\!-\!1} Z_0(n_1\!+\!n_2\!-\!k\!-\!2) Z_0(D\!+\!k\!-\!5)
\label{eq:ZddinstemThree2} \\
& -q^4 \xi^2 Z_0(D\!-\!3) \sum_{k=3}^{n_2\!-\!1} Z_0(n_1\!+\!n_2\!-\!k\!-\!2) Z_0(k\!-\!3)
\label{eq:ZddinstemThree3} \\
& + O((1-\xi)), \notag
\end{align}
	\label{eq:ZddinstemThree}
\end{subequations}
where we have used the zeroth order expansions
Eqs.~(\ref{eq:Ztwoseg_simple}) and~(\ref{eq:Cd_with_Z0_final}) of $\Ztwoseg$
and $\Zdinstem$, respectively, in the second equality.

At this point, all partition functions for structures in
$\Zddinstem{1}{1}-\Zddtwostem-\ZddinstemMutual=\ZddinstemOne+\ZddinstemTwo+\ZddinstemThree$
have been written down to the zeroth order in $(1-\xi)$ in
Eqs. (\ref{eq:ZddinstemOne}), (\ref{eq:ZddinstemTwo}), and
(\ref{eq:ZddinstemThree}). The next task is then summing over all the
terms in the three equations (a total of eleven terms, four in each of
Eqs. (\ref{eq:ZddinstemOne}) and (\ref{eq:ZddinstemTwo}), and three in
Eq. (\ref{eq:ZddinstemThree})).  This task is going to be accomplished
by the following steps.  First, the eleven terms will be
combined into several subgroups and the summations in each of the
subgroups will be evaluated and simplified separately.  Finally, these
results  will be combined together into a final expression for
$\Zddinstem{1}{1}-\Zddtwostem-\ZddinstemMutual$.

The first of these subgroups includes the terms
~(\ref{eq:ZddinstemOne1}), (\ref{eq:ZddinstemTwo1}), and
(\ref{eq:ZddinstemThree1}).  In order to combine these terms,
we apply changes of summation variable to the summation in
(\ref{eq:ZddinstemTwo1})
\begin{displaymath}
\begin{aligned}
& q^3 \xi \sum_{k=2}^{D-2} Z_0(k-2) Z_0(n_1+n_2+D-k-4) \\
= & q^3 \xi \!\!\!\!\!\!\sum_{k'=n_1+n_2+D-4}^{n_1+n_2}\!\!\!\!\!\! Z_0(n_1+n_2+D-k'-4)Z_0(k'-2)
\end{aligned}
\end{displaymath}
and to the summation in (\ref{eq:ZddinstemThree1})
\begin{displaymath}
\begin{aligned}
& q^3 \xi  \sum_{k=2}^{n_2-1} Z_0(n_1+n_2-k-2) Z_0(D+k-4) \\
= & q^3 \xi  \!\!\!\!\!\!\sum_{k'=n_1+n_2-2}^{n_1+1}\!\!\!\!\!\! Z_0(k'-2)Z_0(n_1+n_2+D-k'-4),
\end{aligned}
\end{displaymath}
which gives them the same form as the summation in (\ref{eq:ZddinstemOne1}).
Thus, these three terms can be combined to
\begin{equation}
\begin{aligned}
&q^3 \xi \left( \sum_{k=2}^{n_1-1} + \sum_{k=n_1+1}^{n_1+n_2-2} + \sum_{k = n_1+n_2 }^{n_1+n_2+D-4} \right) Z_0(k-2) \times \\
& \qquad\qquad\qquad\qquad\qquad \qquad Z_0(n_1+n_2+D-4-k) \\
& = q^2 \sum_{k=2}^{n_1+n_2+D-4} Z_b(k) Z_0(n_1+n_2+D-4-k) \\
& \quad -q^3 \xi \left[Z_0(n_1-2)Z_0(n_2+D-4) \right. \\
& \qquad\qquad\qquad\qquad + \left. Z_0(n_1+n_2-3) Z_0(D-3)\right] \\
& \quad + O((1-\xi)) \\
& = q^2 \left[ Z_0(n_1+n_2+D-4) -Z_0(n_1+n_2+D-5) \right] \\
& \quad -q^3 \xi \left[ Z_0(n_1-2)Z_0(n_2+D-4) \right. \\
& \qquad\qquad\qquad\qquad + \left. Z_0(n_1+n_2-3) Z_0(D-3)\right] \\
& \quad + O((1-\xi)),
\end{aligned}
\label{eq:subgroup1}
\end{equation}
where we have used Eq.~(\ref{eq:Zb-to-Z0}) in the first equality to
replace $q\xi Z_0(k-2)$ by $Z_b(k)$ up to terms of order $(1-\xi)$ and
Eq.~(\ref{eq:cyclicRNA_view}) in the second equality to express the
summation as a simple combination of partition functions.

The second subgroup comprises the terms~(\ref{eq:ZddinstemOne2}),
(\ref{eq:ZddinstemTwo2}), and~(\ref{eq:ZddinstemThree2}).  Again,
we apply a change of summation variable to the term~(\ref{eq:ZddinstemTwo2})
\begin{displaymath}
\begin{aligned}
& -q^3 \xi \sum_{k=2}^{D-2} Z_0(k-2) Z_0(n_1+n_2+D-k-5) \\
= & -q^3 \xi \!\!\!\!\!\!\sum_{k'=n_1+n_2+D-5}^{n_1+n_2-1}\!\!\!\!\!\!\! Z_0(n_1+n_2+D-k'-5)Z_0(k'-2)
\end{aligned}
\end{displaymath}
and to the term~(\ref{eq:ZddinstemThree2})
\begin{displaymath}
\begin{aligned}
&-q^3 \xi  \sum_{k=2}^{n_2-1} Z_0(n_1+n_2-k-2) Z_0(D+k-5) \\
= & -q^3 \xi\!\!\!\!\!\!  \sum_{k'=n_1+n_2-2}^{n_1+1}\!\!\!\!\!\! Z_0(k'-2)Z_0(n_1+n_2+D-k'-5)
\end{aligned}
\end{displaymath}
such that again all three terms in the subgroup have the same form.
Then, their combination can be simplified as
\begin{equation}
\begin{aligned}
&-q^3 \xi \left( \sum_{k=2}^{n_1-1} + \sum_{k=n_1+1}^{n_1+n_2-2} + \sum_{k = n_1+n_2-1 }^{n_1+n_2+D-5} \right) Z_0(k-2) \times \\
& \qquad\qquad\qquad\qquad\qquad \qquad Z_0(n_1+n_2+D-5-k) \\
& = -q^2 \sum_{k=2}^{n_1+n_2+D-5} Z_b(k) Z_0(n_1+n_2+D-5-k) \\
& \quad +q^3 \xi Z_0(n_1-2)Z_0(n_2+D-5) \\
& \quad + O((1-\xi)) \\
& = -q^2 \left[ Z_0(n_1+n_2+D-5) -Z_0(n_1+n_2+D-6) \right] \\
& \quad +q^3 \xi Z_0(n_1-2)Z_0(n_2+D-5) \\
& \quad + O((1-\xi))\\
\end{aligned}
\label{eq:subgroup2}
\end{equation}
using the same relations as above.

The third subgroup combines the terms~(\ref{eq:ZddinstemOne3})
and~(\ref{eq:ZddinstemTwo3}).  Upon applying the change of variables
\begin{displaymath}
\begin{aligned}
&  -q^4 \xi^2 Z_0(n_2\!-\!2) \sum_{k = 2}^{D-2} Z_0(k\!-\!2) Z_0(n_1\!+\!D\!-\!k\!-\!4) \\
= & -q^4 \xi^2 Z_0(n_2\!-\!2) \sum_{k' =n_1\!+\!D\!-\!4}^{n_1}  Z_0(n_1\!+\!D\!-\!k'\!-\!4) Z_0(k'\!-\!2)
\end{aligned}
\end{displaymath}
to the term~(\ref{eq:ZddinstemTwo3}) it takes the same form as the
term~(\ref{eq:ZddinstemOne3}) such that their combination can be
simplified to
\begin{equation}
\begin{aligned}
& -q^4 \xi^2 Z_0(n_2-2) \left( \sum_{k=2}^{n_1-1} + \sum_{k =n_1}^{n_1+D-4}  \right) Z_0(k-2) \times \\
& \qquad\qquad\qquad\qquad\qquad \qquad  Z_0(n_1+D-k-4) \\
& = -q^3 \xi Z_0(n_2-2)  \sum_{k=2}^{n_1+D-4} Z_b(k)Z_0(n_1+D-4-k) \\
& \quad + O((1-\xi)) \\
& =  -q^3 \xi Z_0(n_2-2) \left[ Z_0(n_1+D-4) - Z_0(n_1+D-5)\right] \\
& \quad + O((1-\xi)).
\end{aligned}
\label{eq:subgroup3}
\end{equation}
The fourth subgroup comprising terms~(\ref{eq:ZddinstemOne4})
and~(\ref{eq:ZddinstemThree3}) is similarly simplified through the
change of summation variable
\begin{displaymath}
\begin{aligned}
& -q^4 \xi^2 Z_0(D-3) \sum_{k=3}^{n_2-1}  Z_0(n_1\!+\!n_2\!-\!k\!-\!2) Z_0(k\!-\!3) \\
= & -q^4 \xi^2 Z_0(D-3)\!\!\! \sum_{k' = n_1\!+\!n_2\!-\!3}^{n_1+1} \!\!\! Z_0(k'\!-\!2)Z_0(n_1\!+\!n_2\!-\!k'\!-\!3),  
\end{aligned}
\end{displaymath}
applied to the term~(\ref{eq:ZddinstemThree3}) which renders it of the
same form as the term~(\ref{eq:ZddinstemOne4}) and allows their
combination into
\begin{equation}
\begin{aligned}
&-q^4 \xi^2 Z_0(D-3) \left( \sum_{k=2}^{n_1-2} + \sum_{k = n_1+1}^{n_1+n_2-3} \right) Z_0(k-2) \times \\
& \qquad\qquad\qquad\qquad\qquad \qquad Z_0(n_1+n_2-3-k) \\
& = -q^3 \xi Z_0(D-3) \sum_{k=2}^{n_1+n_2-3} Z_b(k)Z_0(n_1+n_2-3-k) \\
& \quad + q^4 \xi^2 Z_0(D-3)\left[ Z_0(n_1-3)Z_0(n_2-2) \right. \\
&  \qquad\qquad\qquad\qquad\qquad\qquad + \left. Z_0(n_1-2)Z_0(n_2-3)\right] \\
& \quad + O((1-\xi)) \\
& = -q^3 \xi Z_0(D-3) \left[ Z_0(n_1+n_2-3) - Z_0(n_1+n_2-4) \right] \\
& \quad + q^4 \xi^2 Z_0(D-3)\left[ Z_0(n_1-3)Z_0(n_2-2) \right. \\
&  \qquad\qquad\qquad\qquad\qquad\qquad + \left. Z_0(n_1-2)Z_0(n_2-3)\right] \\
& \quad + O((1-\xi)).
\end{aligned}
\label{eq:subgroup4}
\end{equation}

The last of the eleven terms is term~(\ref{eq:ZddinstemTwo4}).
Using the same relations Eq.~(\ref{eq:Zb-to-Z0}) and
Eq.~(\ref{eq:cyclicRNA_view}) as before, this term can be evaluated
by itself as follows:
\begin{equation}
\begin{aligned}
& -q^4 \xi^2 Z_0(n_1+n_2-3)  \sum_{k=2}^{D-3} Z_0(k-2) Z_0(D-k-3) \\
& = -q^3 \xi Z_0(n_1+n_2-3) \sum_{k=2}^{D-3} Z_b(k) Z_0(D-k-3) \\
& \quad + O((1-\xi)) \\
& = -q^3 \xi Z_0(n_1+n_2-3)\left[ Z_0(D-3)-Z_0(D-4)\right] \\
& \quad + O((1-\xi)).
\end{aligned}
\label{eq:subgroup5}
\end{equation}

Collecting all five subgroups Eqs.~(\ref{eq:subgroup1}),
(\ref{eq:subgroup2}), (\ref{eq:subgroup3}), (\ref{eq:subgroup4}),
and~(\ref{eq:subgroup5}) and adding them to the zeroth order
expressions Eqs.~(\ref{eq:Zddtwostem_0th-approx})
and~(\ref{eq:ZddinstemMutual}) for $\Zddtwostem$ and
$\ZddinstemMutual$, respectively, we finally obtain
\begin{widetext}
\begin{equation}
\begin{aligned}
\Zddinstem{1}{1} & = q^2 \xi Z_0(D-2)Z_0(n_1+n_2-2) \\
& - q^3 \xi Z_0(D-3) Z_0(n_1+n_2-3) \\
& + q^2 \left[ Z_0(n_1+D+n_2-4) - 2 Z_0(n_1+D+n_2-5) + Z_0(n_1+D+n_2-6) \right] \\
& -q^3 \xi \left\{ Z_0(n_1-2)\left[ Z_0(n_2+D-4) -Z_0(n_2+D-5) \right] \right. \\
& \left. \qquad \qquad + Z_0(n_2-2)\left[Z_0(n_1+D-4)-Z_0(n_1+D-5) \right] \right\} \\
& + q^3 \xi \left[Z_0(D-3)Z_0(n_1+n_2-4)+Z_0(D-4)Z_0(n_1+n_2-3)  \right] \\
& + q^4 \xi^2 Z_0(D-3) \left[ Z_0(n_1-3)Z_0(n_2-2) +  Z_0(n_1-2)Z_0(n_2-3)  \right] \\
& + O((1-\xi)).
\end{aligned}
	\label{eq:P11_final}
\end{equation}
\end{widetext}

In the limit $N/2\approx n_1\approx n_2 \gg D \gg 1$, we can insert
the asymptotic form Eq.~(\ref{eq:G_BC}) for $Z_0$.  As before,
any term in which $n_1$ and $n_2$ occur as arguments of different $Z_0$'s
depend on $N$ as $N^{-3}$ and can thus be neglected compared to the
terms in which $n_1+n_2$ is the argument of one $Z_0$ such that only
the terms in the first, second, third, and sixth line contribute in
the limit of large $N$.  Inserting the asymptotic form Eq.~(\ref{eq:G_BC})
for all $Z_0$ in these lines finally yields
\begin{equation}
\begin{aligned}
\Zddinstem{1}{1} = & \frac{q^2}{z^6}A\frac{(z-1)^2}{N^{3/2}}z^{n_1+D+n_2}\\
& + \left(\frac{q^2\xi}{z^4}- \frac{q^3\xi}{z^6}+ \frac{2q^3\xi}{z^7}\right)
 A^2\frac{1}{D^{3/2}N^{3/2}}z^{n_1+D+n_2}\\
& + O((1-\xi),z^{n_1+D+n_2}D^{-5/2},z^{n_1+D+n_2}N^{-5/2}).
\end{aligned}
	\label{eq:Cdd_final}
\end{equation}
We note that the first line of this result is precisely the asymptotic
expansion of the naive expectation
Eq.~(\ref{eq:Zddinstem11_naive_ini}) while the second line has the
same power law dependence on the distance $D$ between the protein
binding sites as the first order term of the correlation function
$g(D)$ calculated above.

\subsection{Numerator of the correlation function}

The numerator of the correlation function $g(D)$ is given by
Eq.~(\ref{eq:gD_numerator}).  We have already argued in
Appendix~\ref{app:gD_calculation} that the terms on the first line
vanish in the limit of large $N$ through an argument that was
independent of the expansion in $(1-\xi)$.  We also argued that the
terms in the second and third line vanish in the limit of large $N$ to
first order in $(1-\xi)$.  In principle, the terms on the second and
third line could still yield a second order contribution in $(1-\xi)$.
However, using the first order expansion
Eq.~(\ref{eq:Zdinstem1st_final}) of $\Zdinstem$, we find that the
first order term of the asymptotic form of $\Zdinstem(n,m)$
is just $q/z^3$ times the zeroth order term.  Therefore, the
contributions of the differences in the second and third line
to the second order in $(1-\xi)$ must vanish in the limit $N \to \infty$
as well.

It is then clear that only the last two terms can contribute to the
numerator ${\cal N}$ of the correlation function $g(D)$ to second
order in $(1-\xi)$. Thus, the numerator of $g(D)$ in the limit of $N
\gtrsim n_1 \approx n_2 \gg D \gg l \geq 1$ becomes
\begin{equation}
\begin{aligned}
& \frac{Z_{dd}}{Z_0} - \frac{Z_d}{Z_0} \times \frac{Z_d}{Z_0} \\
& =  \xi (1-\xi) \frac{ \Zddtwostem}{Z_0(N)}\! +\! (1\!-\!\xi)^2\! \left[ \frac{ \Zddinstem{1}{1}}{Z_0(N)} - \frac{\Zdinstem}{Z_0(N)}\! \times\! \frac{\Zdinstem}{Z_0(N)} \right]\\
&\quad +O((1-\xi)^3,N^{-1}), 
\end{aligned}
	\label{eq:gD_large_simplified}
\end{equation}
in which the leading term is $O((1-\xi))$.  Dividing the asymptotic
expressions for $\Zdinstem$, $\Zddtwostem$, and $\Zddinstem{1}{1}$
calculated above (Eqs.~(\ref{eq:Cd_final}),
(\ref{eq:Zddtwostem_first_final}), and~(\ref{eq:Cdd_final}),
respectively) by the asymptotic expression Eq.~(\ref{eq:G_BC}) for
$Z_0$ yields
\begin{equation}
\begin{aligned}
\frac{\Zddtwostem(n_1,D,n_2)}{Z_0(N)} = & \frac{1}{z^{2l}}\left( \frac{q^2 \xi }{z^4} + (1\!-\!\xi)\frac{q^3 \xi(z-1)}{z^7} \right)  \frac{A}{D^{3/2}} \\
& + O((1-\xi)^2,N^{-1},D^{-5/2}) \\
\frac{\Zddinstem{1}{1}(n_1,D,n_2)}{Z_0(N)} = & \frac{q^2 (z-1)^2}{z^{(6+2l)}} \\
&  + \frac{1}{z^{2l}}\left( \frac{q^2 \xi}{z^4} - \frac{q^3 \xi(z-2)}{z^7} \right) \frac{A}{D^{3/2}} \\
& + O((1-\xi),N^{-1},D^{-5/2})\\
\frac{\Zdinstem(n_1,D\!+\!l\!+\!n_2)}{Z_0(N)} = & \frac{\Zdinstem(n_1+l+D, n_2)}{Z_0(N)}  \\
= & \frac{q(z-1)}{z^{3+l}} + O((1-\xi),N^{-1},D^{-5/2}).
\end{aligned}
\end{equation}
in the limit of $N=n_1+D+n_2+2l \gtrsim n_1 \approx n_2 \gg D \gg l
\geq 1$.  Substituting these fractions into
Eq.~(\ref{eq:gD_large_simplified}) shows that the terms independent of
$D$ cancel each other, thus yielding the asymptotic form of the
numerator of $g(D)$ as
\begin{equation}
\begin{aligned}
{\cal N}=&
\frac{A}{z^{2l} D^{3/2}} \left[ (1-\xi)\frac{q^2 \xi^2}{z^4} +
(1-\xi)^2 \left(\frac{q^2 \xi}{z^4} + \frac{q^3\xi}{z^7}\right) \right]\\
&\quad + O((1-\xi)^3,N^{-1},D^{-5/2}).
\end{aligned}
\label{eq:gD_up_1st_final}
\end{equation}


\subsection{Denominator}

The denominator of the correlation function $g(D)$ is given by
Eq.~(\ref{eq:g_down_initial}) and now has to be calculated to first
order in $(1-\xi)$. With the help of Eq.~(\ref{eq:Zd-S-Z0}) and the
asymptotic form Eq.~(\ref{eq:Cd_final}) of $\Zdinstem$, the first
ratio in Eq. (\ref{eq:g_down_initial}) can be rewritten as
\begin{equation}
\begin{aligned}
&  \frac{Z_d(n_1,D+l+n_2 )}{Z_0(N)} \\
& = \frac{Z_0(N-l)}{Z_0(N)} - (1-\xi) \frac{\Zdinstem(n_1,D+l+n_2)}{Z_0(N)} \\
& = \left(1\! -\!  (1\!-\!\xi)\frac{q(z-1)}{z^3} \right) \frac{N^{3/2}}{z^l(N-l)^{3/2}}\! +\! O\left((1\!-\!\xi)^2,\!N^{-1}\right) \\
& = \frac{1}{z^l} \left(1 -  \frac{q(1-\xi)(z-1)}{z^3} \right)
+ O\left((1-\xi)^2,\!N^{-1} \right).
\end{aligned}
\end{equation}
By symmetry, the asymptotic form of the second ratio in
Eq.~(\ref{eq:g_down_initial}) must be the same.
The third ratio, which includes 
$Z_{dd}$, is derived to the first order in $(1-\xi)$ as
\begin{equation}
\begin{aligned}
& \frac{Z_{dd}(n_1,D,n_2)}{Z_0(N)} \\
& = \frac{Z_0(N-2l)}{Z_0(N)} \\
& \quad - (1-\xi)\left[ \frac{\Zdinstem(n_1,D+n_2)}{Z_0(N)} + \frac{\Zdinstem(n_1+D,n_2)}{Z_0(N)} \right] \\
& \quad - (1-\xi)^2 \frac{\Zddinstem{1}{1}}{Z_0(N)} + \xi(1-\xi) \frac{\Zddtwostem}{Z_0(N)}  \\
& = \frac{1}{z^{2l}} \left[ 1- \frac{2q(1-\xi)(z-1)}{z^3} + \frac{q^2 \xi^2 (1-\xi)}{z^4} \frac{A}{D^{3/2}} \right] \\
& \quad + O\left( (1-\xi)^2,N^{-1},D^{-5/2}\right)
\end{aligned}
\end{equation}
using Eq.~(\ref{eq:Zdd_final}) in the first equality and the
asymptotic expressions Eqs.~(\ref{eq:Cd_final})
and~(\ref{eq:Zddtwostem_0th_asympt}) for $\Zdinstem$ and
$\Zddtwostem$, respectively, in the second equality.

Combining all four terms we find
\begin{equation}
\begin{aligned}
{\cal D}=& \left(1 + \frac{c_1}{K_{d,1}^{(0)}z^l}\right)
\left( 1 + \frac{c_2}{K_{d,2}^{(0)}z^l}\right)\\
&-(1-\xi)\left[
\frac{q(z-1)}{z^{3+l}} \left( \frac{c_1}{K_{d,1}^{(0)}} +  \frac{c_2}{K_{d,2}^{(0)}} \right)\right.\\
&\qquad\qquad +\left. \frac{2q(z-1)}{z^{3+2l}} \frac{c_1 c_2}{K_{d,1}^{(0)}K_{d,2}^{(0)} }\right]\\
&\quad +O((1-\xi)^2,N^{-1},D^{-3/2}).
\end{aligned}
	\label{eq:gD_down_1st_final}
\end{equation}
where we ignored the terms depending on the distance $D$ between the
binding sites since they are subleading to the constant term and
we have neglected other subleading terms in the distance $D$ in the
numerator already as well.


\subsection{Correlation function}

Dividing Eq.~(\ref{eq:gD_up_1st_final}) by the square of
Eq.~(\ref{eq:gD_down_1st_final}) yields the correlation function $g(D)$
with the overall shape
\begin{equation}
g(D)=(1-\xi)\frac{{\cal A}}{D^{3/2}}+O((1-\xi)^3,N^{-1},D^{-5/2})
\end{equation}
where ${\cal A}$ is in principle given by an explicit expression
of the parameters $z$ and $A$ of the partition function $Z_0$,
the loop cost $(1-\xi)$ (up to first order), the concentrations $c_1$
and $c_2$ of the proteins, and the bare equilibirum constants
$K_{d,1}^{(0)}$ and $K_{d,2}^{(0)}$ of the two binding sites.  For small
protein concentrations $c_i\ll K_{d,i}^{(0)}z^l$ this prefactor
simplifies to
\begin{equation}
{\cal A}_{\mathrm{low}\,c}=\frac{A}{z^{2l}} \left[\frac{q^2 \xi^2}{z^4} +
(1-\xi) \left(\frac{q^2 \xi}{z^4} + \frac{q^3\xi}{z^7}\right) \right].
\label{eq:A_low_c}
\end{equation}
and has to be evaluated numerically for arbitrary protein concentrations.

\end{document}